\documentclass[12pt]{article}
\usepackage{amsmath}
\usepackage[unicode,bookmarks,bookmarksopen,bookmarksopenlevel=2,colorlinks,linkcolor=blue,citecolor=green]{hyperref}

\def\comment#1{}
\input{tcilatex}

\begin{document}

\title{The Hamiltonian approach in classification and integrability of
hydrodynamic chains}
\author{Maxim V. Pavlov}
\date{}
\maketitle

\begin{abstract}
New approach in classification of integrable hydrodynamic chains is
established. This is the method of the \textit{Hamiltonian} hydrodynamic
reductions. Simultaneously, this approach yields \textit{explicit}
Hamiltonian hydrodynamic reductions of the Hamiltonian hydrodynamic chains.
The concept of \textit{reducible} Poisson brackets is established. Also this
approach is useful for non-Hamiltonian hydrodynamic chains. The \textit{%
deformed} Benney hydrodynamic chain is considered.
\end{abstract}

\tableofcontents

\bigskip \bigskip

\textit{keywords}: Poisson bracket, Hamiltonian structure, hydrodynamic
chain, Liouville coordinates, reciprocal transformation, Miura
transformation, Gibbons equation, Riemann mapping

MSC: 35L40, 35L65, 37K10;\qquad PACS: 02.30.J, 11.10.E.

\bigskip

\section{Introduction}

This paper is devoted to classification and integrability of 2+1 quasilinear
equations and corresponding hydrodynamic chains (see \textbf{\cite{Fer+Dav}}%
, \textbf{\cite{Maks+Egor}}, \textbf{\cite{Maks+eps}}) by the
``Hamiltonian'' approach based on the Hamiltonian formulation of integrable
hydrodynamic chains and their hydrodynamic reductions. Nevertheless, this
Hamiltonian approach is much more universal, because most of explicit
hydrodynamic reductions are non-Hamiltonian and this approach is useful as
well as for non-Hamiltonian hydrodynamic chains. However, without lost of
generality an illustration of this approach will be given on important
examples connected with so-called $M$--Poisson brackets (see details in 
\textbf{\cite{Maks+Puas}}).

A hydrodynamic type system is $N$ component quasilinear system of PDE's of
the first order (see \textbf{\cite{Dubr+Nov}})%
\begin{equation}
u_{t}^{i}=\upsilon _{j}^{i}(\mathbf{u})u_{x}^{j}\text{, \ \ \ }i,j=1,2,...,N.
\label{first}
\end{equation}%
A hydrodynamic chain is a generalization of a hydrodynamic type system on an
infinite component case (see details below). At the same time, a hierarchy
of commuting hydrodynamic chains is equivalent to a family of 2+1 integrable
quasilinear equations (see \textbf{\cite{Fer+Kar}}, \textbf{\cite{Maks+Egor}}%
, \textbf{\cite{Maks+eps}}). Hydrodynamic chains and corresponding 2+1
quasilinear equations can be decomposed in infinitely many ways on families
of 1+1 integrable hydrodynamic type systems (\textbf{\ref{first}}) (see 
\textbf{\cite{Fer+Kar}}, \textbf{\cite{Gib+Tsar}}, \textbf{\cite{Maks+eps}}%
). Nonlinear PDE's describing such hydrodynamic reductions have solutions
parameterized by $N$ arbitrary functions of a single variable (see \textbf{%
\cite{Fer+Kar}}, \textbf{\cite{Gib+Tsar}}). These nonlinear PDE's are still
not solved yet (except \textbf{\cite{Maks+eps}}), but infinitely many
particular \textit{explicit} solutions parameterized by an arbitrary number
of constants been known for the Benney hierarchy many years ago (see \textbf{%
\cite{Gib+Yu}}, \textbf{\cite{Kodama}}, \textbf{\cite{Krich}}, \textbf{\cite%
{KM}}, \textbf{\cite{Zakh}}). Moreover, new hydrodynamic reductions can be
found by this Hamiltonian approach (see, for instance, \textbf{\cite%
{Maks+Kuper} }and \textbf{\cite{Maks+wdvv}}). Every hydrodynamic reduction
determines some particular solution of corresponding 2+1 quasilinear
equation. If such an equation is homogeneous, then corresponding solution is
self-similar. Such solutions can be found by different methods (see \textbf{%
\cite{Manas}}), based on the twistor approach, theory of the Beltrami and
the Hamilton-Jacobi equations, the conformal mapping and the dispersionless $%
\overline{\partial }-$method, differential and algebraic geometry (see also 
\textbf{\cite{Heavenly}}). An application of the generalized hodograph
method yields most general non-self similar solutions by the nonlinear
superposition principle (see \textbf{\cite{Tsar}}). For instance, the
``symmetry constraint'' method established in \textbf{\cite{Bogdan}} comes
from a theory of 2+1 integrable dispersive nonlinear PDE's such as the
Kadomtsev--Petviashvili (KP) or the Veselov--Novikov (VN) equations. The
Hamiltonian approach is much more \textit{universal}, because this method
exists \textit{irrespective of origin} of given hydrodynamic chain.

The problem of classification of Hamiltonian chains consists of two steps.
This Hamiltonian approach is used to suggest some list of new Poisson
brackets connected with hydrodynamic chains. Such classification of the
Poisson brackets should be done in further publications (see also \textbf{%
\cite{Maks+Puas}}), at least all ``reducible'' Poisson brackets (see
definition below) should be described (nevertheless, we believe that any
Poisson bracket is reducible by some appropriate ``moment decomposition'').
Such \textit{linear} Poisson brackets have been introduced by I. Dorfman in 
\textbf{\cite{Dorfman}}. The \textbf{main claim} of this paper is that 
\textit{the Hamiltonian structure completely determines hierarchy of
integrable hydrodynamic chains up to the Miura type and reciprocal
transformations} (This is not a true for integrable dispersive and
hydrodynamic type systems). Thus, a description of the Hamiltonian
structures means a description of integrable hydrodynamic chains. In this
paper without lost of generality we restrict our consideration on the
Hamiltonian densities depending on first three moments $A^{k}$. Obviously,
they can be \textit{linear }$\mathbf{H}=A^{2}+f(A^{0},A^{1})$, \textit{%
quasilinear} $\mathbf{H}=g(A^{0},A^{1})A^{2}+f(A^{0},A^{1})$ and \textit{%
fully nonlinear} $\mathbf{H}(A^{0},A^{1},A^{2})$. Since the integrable
hydrodynamic chains determined by the \textit{quasilinear} and \textit{fully
nonlinear} Hamiltonian densities can be reduced (by some appropriate
reciprocal transformations) to the integrable hydrodynamic chains determined
by the \textit{linear} Hamiltonian densities, we restrict our consideration
on the \textit{linear} case in this paper only. Two others case are just
briefly described.

The paper is organized in the following order. In the second section the
method allowing to extract explicit Hamiltonian hydrodynamic reductions with
the aid of pseudo-potentials is established. In the third section several 
\textit{reducible} multi-parametric families of the Poisson brackets are
derived. In the fourth section the integrability criteria of hydrodynamic
chains are considered. In the fifth section the first such a family of the
Poisson brackets is investigated by three different methods. Corresponding
integrable hydrodynamic chains are found by the requirement of the existence
of extra conservation law creating a commuting flow (see \textbf{\cite{Kuper}%
}, \textbf{\cite{Maks+Egor}}), by the method of pseudopotentials (see 
\textbf{\cite{Fer+Kar}}, \textbf{\cite{Zakh+multi}}) and separately by an
application of the \textit{phenomenological} Hamiltonian approach. All these
hydrodynamic chains are connected with integrable hydrodynamic chains from
the sixth section by a reciprocal transformation. In the sixth section
hydrodynamic chains associated with the Kupershmidt Poisson brackets are
considered. In the seventh section a classification of integrable
Hamiltonian hydrodynamic chains associated with the Kupershmidt--Manin
Poisson bracket is significantly simplified for the \textit{linear}
Hamiltonian density (cf. \textbf{\cite{Fer+Dav}}). In the eighth section the
method of Hamiltonian hydrodynamic reductions is extended on non-Hamiltonian
hydrodynamic chains. Corresponding hydrodynamic type systems are
non-Hamiltonian too. In the ninth section the modified Kupershmidt
hydrodynamic chain is investigated with the aid of the Zakharov hydrodynamic
reductions.

\section{The Hamiltonian approach for 2+1 quasilinear equations}

The ``Hamiltonian'' approach is universal and based on the ``concept'' of
pseudo--potentials (see \textbf{\cite{Fer+Kar}}, \textbf{\cite{Maks+Egor}}, 
\textbf{\cite{Zakh+multi}}). Let us briefly describe and illustrate this
approach on the famous Benney hydrodynamic chain \textbf{\cite{Benney}}%
\begin{equation}
A_{t}^{k}=A_{x}^{k+1}+kA^{k-1}A_{x}^{0}\text{, \ \ \ \ \ }k=0,1,2,...
\label{bm}
\end{equation}%
and corresponding the Khohlov--Zabolotzkaya equation%
\begin{equation*}
(A_{y}^{0}-A^{0}A_{x}^{0})_{x}=A_{tt}^{0},
\end{equation*}%
which can be obtained with the aid of two first equations from the Benney
hydrodynamic chain and the first equation from its first commuting flow%
\begin{equation*}
A_{y}^{k}=A_{x}^{k+2}+A^{0}A_{x}^{k}+(k+1)A^{k}A_{x}^{0}+kA^{k-1}A_{x}^{1}%
\text{, \ \ \ \ \ }k=0,1,2,...
\end{equation*}%
by eliminating of the moments $A^{1}$ and $A^{2}$. The method of
hydrodynamic reductions developed for these equations (see \textbf{\cite%
{Gib+Tsar}}) was extended for more wide class of 2+1 hydrodynamic type
systems (see \textbf{\cite{Fer+Kar}}), more general 2+1 quasilinear
equations and hydrodynamic chains (see \textbf{\cite{FerKarMax}} and \textbf{%
\cite{FKT}}) as an integrability criterion. The above equations have $N$
component hydrodynamic reductions written in the Riemann invariants%
\begin{equation}
r_{t}^{i}=\mu ^{i}(\mathbf{r})r_{x}^{i}\text{, \ \ \ \ \ \ \ \ \ }%
r_{y}^{i}=(\mu ^{i^{2}}(\mathbf{r})+A^{0}(\mathbf{r}))r_{x}^{i}\text{, \ \ \
\ \ \ \ }i=1,2,...,N  \label{redu}
\end{equation}%
consistent with the generating functions of conservation laws%
\begin{equation}
\mu _{t}=\partial _{x}(\frac{\mu ^{2}}{2}+A^{0})\text{, \ \ \ \ \ \ \ }\mu
_{y}=\partial _{x}(\frac{\mu ^{3}}{3}+A^{0}\mu +A^{1}),  \label{gen}
\end{equation}%
where all moments $A^{k}$ are some functions of the Riemann invariants $%
r^{n} $. Following \textbf{\cite{Gib+Tsar}}, the Gibbons--Tsarev system%
\begin{equation}
\partial _{i}\mu ^{k}=\frac{\partial _{i}A^{0}}{\mu ^{i}-\mu ^{k}}\text{, \
\ \ }\partial _{ik}A^{0}=2\frac{\partial _{i}A^{0}\partial _{k}A^{0}}{(\mu
^{i}-\mu ^{k})^{2}}\text{, \ \ \ \ \ }i\neq k  \label{gt}
\end{equation}%
can be obtained from the compatibility conditions $\partial _{k}(\partial
_{i}\mu )=\partial _{i}(\partial _{k}\mu )$, where%
\begin{equation}
\partial _{i}\mu =\frac{\partial _{i}A^{0}}{\mu ^{i}-\mu },  \label{tri}
\end{equation}%
which is a consequence that $\mu $ is a conservation law density of the
hydrodynamic type systems (\textbf{\ref{redu}}). Taking into account all
moments $A^{k}$ are connected with the Benney hydrodynamic chain by the
Laurent series%
\begin{equation}
\lambda =\mu +\frac{A^{0}}{\mu }+\frac{A^{1}}{\mu ^{2}}+\frac{A^{2}}{\mu ^{3}%
}+...  \label{ryad}
\end{equation}%
one can verify that the Benney hydrodynamic chain together with any its
hydrodynamic reductions satisfy the Gibbons equation (see \textbf{\cite%
{Gibbons}})%
\begin{equation}
\lambda _{t}-\mu \lambda _{x}=\frac{\partial \lambda }{\partial \mu }[\mu
_{t}-\partial _{x}(\frac{\mu ^{2}}{2}+A^{0})].  \label{gib}
\end{equation}%
We believe that the Gibbons--Tsarev system is Darboux integrable, because
plenty particular solutions parameterized by arbitrary constants are known.
However, a general solution still is unknown.

In this paper the Hamiltonian approach is established. Let us choose $N$
arbitrary conservation law densities $\mu ^{(k)}=a^{k}$, then%
\begin{equation}
a_{t}^{k}=\partial _{x}\left( \frac{(a^{k})^{2}}{2}+A^{0}(\mathbf{u})\right)
.  \label{for}
\end{equation}%
The transformation $a^{k}(\mathbf{r})$ is invertible. Then the
Gibbons--Tsarev system (in the independent variables $a^{k}$ on sole
function $A^{0}$ only, cf. (\textbf{\ref{gt}}))%
\begin{eqnarray}
(a^{i}-a^{k})\partial _{ik}A^{0} &=&\partial _{k}A^{0}\partial _{i}\left(
\sum \partial _{n}A^{0}\right) -\partial _{i}A^{0}\partial _{k}\left( \sum
\partial _{n}A^{0}\right) \text{, \ \ }i\neq k,  \notag \\
&&  \label{egt}
\end{eqnarray}%
\begin{equation*}
(a^{i}-a^{k})\frac{\partial _{ik}A^{0}}{\partial _{i}A^{0}\partial _{k}A^{0}}%
+(a^{k}-a^{j})\frac{\partial _{jk}A^{0}}{\partial _{j}A^{0}\partial _{k}A^{0}%
}+(a^{j}-a^{i})\frac{\partial _{ij}A^{0}}{\partial _{i}A^{0}\partial
_{j}A^{0}}=0\text{, \ \ }i\neq j\neq k,
\end{equation*}%
can be derived from the compatibility conditions (cf. (\textbf{\ref{tri}})) $%
\partial _{k}(\partial _{i}\mu )=\partial _{i}(\partial _{k}\mu )$, where $%
\partial _{i}\equiv \partial /\partial a^{i}$ and%
\begin{equation}
\partial _{i}\mu =\frac{\partial _{i}A^{0}}{\mu -a^{i}}\left[ \sum \frac{%
\partial _{m}A^{0}}{\mu -a^{m}}-1\right] ^{-1}.  \label{ah}
\end{equation}

Assume $a^{k}$ are flat coordinates (see \textbf{\cite{Dubr+Nov}}), then the
corresponding Poisson bracket can be reduced (by linear transformation of
field variables $a^{k}$) to the diagonal form%
\begin{equation}
\{a^{i},a^{j}\}=(\varepsilon _{i})^{-1}\delta ^{ij}\delta ^{\prime
}(x-x^{\prime }),  \label{tot}
\end{equation}%
where $\delta ^{ij}$ is the Kronecker symbol.

\textbf{Theorem 1}: \textit{The hydrodynamic type system} (\textbf{\ref{for}}%
) \textit{with the local Hamiltonian structure}%
\begin{equation}
a_{t}^{k}=\frac{1}{\varepsilon _{k}}\partial _{x}\frac{\partial \mathbf{h}%
_{2}}{\partial a^{k}}  \label{ham}
\end{equation}%
\textit{is a hydrodynamic reduction of the Benney hydrodynamic chain iff} $%
A^{0}=\Sigma \varepsilon _{k}a^{k}$ \textit{and the Hamiltonian density} $%
\mathbf{h}_{2}=\Sigma \varepsilon _{k}(a^{k})^{3}/6+(A^{0})^{2}/2$.

\textbf{Proof}: Let us substitute the expression $A^{0}=\Sigma \varepsilon
_{k}a^{k}$ in (\textbf{\ref{ah}}). This ODE system%
\begin{equation*}
\partial _{i}\mu =\frac{\varepsilon _{i}}{\mu -a^{i}}\left[ \sum \frac{%
\varepsilon _{m}}{\mu -a^{m}}-1\right] ^{-1}
\end{equation*}%
can be integrated in the implicit form%
\begin{equation}
\lambda =\mu -\sum \varepsilon _{k}\ln (\mu -a^{k}),  \label{water}
\end{equation}%
where $\lambda $ is an integration factor. Simultaneously, the equation of
the Riemann surface satisfies the Gibbons equation (\textbf{\ref{gib}}).

\textbf{Corollary}: Since $A^{0}=\Sigma \varepsilon _{k}a^{k}$, then all
other moments%
\begin{equation}
A^{n}=\frac{1}{n+1}\sum \varepsilon _{k}(a^{k})^{n+1},  \label{log}
\end{equation}%
if $\Sigma \varepsilon _{k}=0$. It is easy to prove by comparison (\textbf{%
\ref{bm}}) with (\textbf{\ref{for}}). Then the Laurent series (\textbf{\ref%
{ryad}}) reduces to so-called ``waterbag'' reduction (see \textbf{\cite%
{Gib+Yu}}) for the equation of the Riemann surface (\textbf{\ref{water}}).
If $\Sigma \varepsilon _{k}\neq 0$ the above formula is valid (see \textbf{%
\cite{Kod+water}}), but expressions of the higher moment $A^{n}$ can be
obtained by the substitution of the Laurent series (\textbf{\ref{ryad}})
into the ``deformed'' above equation%
\begin{equation*}
\lambda -\sum \varepsilon _{k}\ln \lambda =\mu -\sum \varepsilon _{k}\ln
(\mu -a^{k}),
\end{equation*}%
because the function $\lambda $ can be replaced by an arbitrary function $%
\tilde{\lambda}=\tilde{\lambda}(\lambda )$ (see (\textbf{\ref{gib}})).

\textbf{Remark}: The famous Zakharov reduction (see \textbf{\cite{Zakh}})%
\begin{equation}
a_{t}^{k}=\partial _{x}\left( \frac{(a^{k})^{2}}{2}+A^{0}\right) \text{, \ \
\ \ \ }b_{t}^{k}=\partial _{x}(u^{k}b^{k})\text{, \ \ \ \ \ }A^{0}=\sum b^{m}
\label{sw}
\end{equation}%
can be obtained in the same way. In comparison with the above ``waterbag''
case, let us expand the Taylor series at the vicinity of the local parameter 
$\lambda $%
\begin{equation*}
\mu ^{(i)}=a^{i}+\lambda b^{i}+\lambda ^{2}c^{i}+...
\end{equation*}%
If first $2N$ conservation law densities $a^{i}$ and $b^{i}$ are
independent, then the above hydrodynamic reduction (without the fixation $%
A^{0}=\Sigma b^{n}$) is given by the moment decomposition%
\begin{equation}
A^{k}=\sum (a^{i})^{k}b^{i}.  \label{rac}
\end{equation}%
Suppose $a^{i}$ and $b^{i}$ are flat coordinates, then a simplest local
Hamiltonian structure is%
\begin{equation}
a_{t}^{k}=\partial _{x}\frac{\partial \mathbf{h}_{2}}{\partial b^{k}}\text{,
\ \ \ \ \ \ }b_{t}^{k}=\partial _{x}\frac{\partial \mathbf{h}_{2}}{\partial
a^{k}},  \label{swh}
\end{equation}%
then the Hamiltonian density $\mathbf{h}_{2}=\Sigma
(a^{n})^{2}b^{n}/2+(A^{0})^{2}/2$, where, indeed, $A^{0}=\Sigma b^{n}$.

For the first time the concept of ``pseudopotentials'' was introduced for
2+1 quasilinear equation obtained by the dispersionless limit of 2+1
dispersive systems or obtained by the longwave (continuum) limit of 2+1
discrete systems in \textbf{\cite{Zakh+multi}}. The meaning of the
``pseudopotential'' is so-called ``dispersionless limit of the Lax
formulation'' for the integrable dispersive systems. This is nothing but the
generating function of conservation laws for hydrodynamic chains (see the
above example (\textbf{\ref{gen}})). Pseudopotentials can be effectively
found for aforementioned 2+1 hydrodynamic type systems (see \textbf{\cite%
{Fer+Kar}}), for 2+1 quasilinear equations (see \textbf{\cite{FerKarMax}})
and for hydrodynamic chains (see \textbf{\cite{Maks+Puas}}). Thus,

\textbf{1}. we suppose pseudopotentials for the Hamiltonian hydrodynamic
chains are given (cf. (\textbf{\ref{gen}})) in the explicit form%
\begin{equation*}
\mu _{t}=\partial _{x}f(\mu ;U^{1},U^{2},...,U^{M}),
\end{equation*}%
where the functions $U^{k}$ depend on a finite number of first moments $%
A^{k} $;

\textbf{2}. we look for explicit hydrodynamic reductions in the Hamiltonian
form (\textbf{\ref{ham}}), where%
\begin{equation*}
d\mathbf{h}=\underset{k=1}{\overset{N}{\sum }}\varepsilon _{k}f(a^{k};U^{1}(%
\mathbf{a}),U^{2}(\mathbf{a}),...,U^{M}(\mathbf{a}))d\mu ^{k}.
\end{equation*}%
The existence of the Hamiltonian density $\mathbf{h}$\ means the existence
of \textit{local} Hamiltonian hydrodynamic reductions. The same procedure
can be used for more general local Hamiltonian structures (see \textbf{\cite%
{Dubr+Nov}}) than (\textbf{\ref{ham}})%
\begin{equation}
a_{t}^{i}=\partial _{x}[\bar{g}^{ik}\frac{\partial \mathbf{h}}{\partial a^{k}%
}]\text{, \ \ \ \ }i=1,2,...,N,  \label{loc}
\end{equation}%
where $\bar{g}^{ij}$ is a constant, non-degenerate and symmetric matrix, and
for \textit{nonlocal} Hamiltonian hydrodynamic reductions (see \textbf{\cite%
{Fer+trans}}).

\textbf{Conjecture 1}: \textit{Any integrable hydrodynamic chain (and
associated 2+1 quasilinear system) admits} $N$ \textit{component
hydrodynamic reductions}%
\begin{equation*}
a_{t}^{k}=\partial _{x}f(a^{k};U^{1}(\mathbf{a}),U^{2}(\mathbf{a}),...,U^{M}(%
\mathbf{a}))
\end{equation*}%
\textit{written in the Hamiltonian form} (\textbf{\ref{loc}}).

Below we reformulate this Hamiltonian approach in the \textit{reverse}
direction: if the hydrodynamic chain has the Hamiltonian structure, then
their explicit hydrodynamic reductions are the Hamiltonian hydrodynamic type
systems.

\section{\textit{Reducible} Poisson brackets and hydrodynamic chains}

The \textbf{main observation} successfully utilized in this paper is that
the first local Hamiltonian structure (\textbf{\ref{swh}}) of the Zakharov
reduction (\textbf{\ref{sw}}) of the Benney hydrodynamic chain (\textbf{\ref%
{bm}}) can be used for a \textit{constructing} the first local Hamiltonian
structure (see \textbf{\cite{KM}})%
\begin{equation}
A_{t}^{k}=\{A^{k},\mathbf{\bar{H}}_{2}\}=[kA^{k+n-1}\partial _{x}+n\partial
_{x}A^{k+n-1}]\frac{\delta \mathbf{\bar{H}}_{2}}{\delta A^{n}},  \label{bhs}
\end{equation}%
where (see (\textbf{\ref{rac}})) the moments $A^{k}=\Sigma (a^{m})^{k}b^{m}$%
. The Kupershmidt--Manin bracket (see (\textbf{\ref{bhs}}))%
\begin{equation}
\{A^{k},A^{n}\}=[kA^{k+n-1}\partial _{x}+n\partial _{x}A^{k+n-1}]\delta
(x-x^{\prime })\text{, \ \ \ \ \ }k,n=0,1,2,...  \label{aca}
\end{equation}%
reduces to the \textit{canonical} Poisson bracket (see (\textbf{\ref{swh}})
and details in \textbf{\cite{Dubr+Nov}}, just non-zero component below)%
\begin{equation}
\{a^{k},b^{k}\}=\delta ^{\prime }(x-x^{\prime })  \label{lg}
\end{equation}%
under the above moment decomposition $A^{k}=\Sigma (a^{m})^{k}b^{m}$. The
Benney hydrodynamic chain (\textbf{\ref{bm}}) is determined by the
Hamiltonian $\mathbf{\bar{H}}_{2}=\frac{1}{2}\int [A^{2}+(A^{0})^{2}]dx$,
while the Kupershmidt--Manin bracket is associated with the momentum $%
\mathbf{\bar{H}}_{1}=\int A^{1}dx$ and the Casimir (annihilator) $\mathbf{%
\bar{H}}_{0}=\int A^{0}dx$. The Zakharov reduction of the Benney
hydrodynamic chain (\textbf{\ref{sw}}) is determined by the Hamiltonian $%
\mathbf{\bar{h}}_{2}=\frac{1}{2}\int [\Sigma (a^{k})^{2}b^{k}+(\Sigma
b^{k})^{2}]dx$, the momentum $\mathbf{\bar{h}}_{1}=\int \Sigma a^{k}b^{k}dx$
and $2N$ Casimirs $\mathbf{\bar{h}}_{(k)}=\int b^{k}dx$, $\mathbf{\bar{h}}%
_{(N+k)}=\int a^{k}dx$.

In general case the local Hamiltonian structure for the hydrodynamic type
system (\textbf{\ref{first}})%
\begin{equation}
u_{t}^{i}=\{u^{i},\mathbf{\bar{h}}\}=[g^{ij}\partial _{x}-g^{is}\Gamma
_{sk}^{j}u_{x}^{k}]\frac{\delta \mathbf{\bar{h}}}{\delta u^{j}},  \label{sec}
\end{equation}%
is determined by the Hamiltonian $\mathbf{\bar{h}}=\int h(\mathbf{u})dx$ and
by the Dubrovin--Novikov bracket (a differential-geometric Poisson bracket
of the first order, see \textbf{\cite{Dubr+Nov}})%
\begin{equation}
\{u^{i}(x),u^{j}(x^{\prime })\}=[g^{ij}\partial _{x}-g^{is}\Gamma
_{sk}^{j}u_{x}^{k}]\delta (x-x^{\prime })\text{, \ \ \ \ \ }i,j=1,2,...,N,
\label{dn}
\end{equation}%
where a \textbf{flat} metric $g^{ij}(\mathbf{u})$ is symmetric and
non-degenerate, $\Gamma _{sk}^{j}=\frac{1}{2}g^{jm}(\partial
_{s}g_{mk}+\partial _{k}g_{ms}-\partial _{m}g_{sk})$ are the Christoffel
symbols. Such local Hamiltonian structure can be written via so-called the 
\textit{Liouville} coordinates $A^{k}(\mathbf{u})$, that the corresponding
Poisson bracket is%
\begin{equation*}
\{A^{k}(x),A^{n}(x^{\prime })\}=[\mathcal{W}^{kn}(\mathbf{A})\partial
_{x}+\partial _{x}\mathcal{W}^{nk}(\mathbf{A})]\delta (x-x^{\prime })\text{,
\ \ \ \ \ }k,n=1,2,...,N.
\end{equation*}%
Formally, this Poisson bracket can be extended on an infinite component
case, see the example (\textbf{\ref{aca}}). \textit{The \textbf{main}
problem in a classification of integrable hydrodynamic chains is a
description of such Poisson brackets}%
\begin{equation}
\{A^{k}(x),A^{n}(x^{\prime })\}=[\mathcal{W}^{kn}(\mathbf{A})\partial
_{x}+\partial _{x}\mathcal{W}^{nk}(\mathbf{A})]\delta (x-x^{\prime })\text{,
\ \ \ \ \ }k,n=1,2,...,  \label{pb}
\end{equation}%
where the coefficients $\mathcal{W}^{kn}(\mathbf{A})$\ satisfy the Jacobi
identity (see \textbf{\cite{Maks+Puas}})%
\begin{eqnarray}
(\mathcal{W}^{ik}+\mathcal{W}^{ki})\partial _{k}\mathcal{W}^{nj} &=&(%
\mathcal{W}^{jk}+\mathcal{W}^{kj})\partial _{k}\mathcal{W}^{ni},  \notag \\
&&  \label{jac} \\
\partial _{n}\mathcal{W}^{ij}\partial _{m}\mathcal{W}^{kn} &=&\partial _{n}%
\mathcal{W}^{kj}\partial _{m}\mathcal{W}^{in}.  \notag
\end{eqnarray}

\textbf{Definition 1}: \textit{A Poisson bracket }(\textbf{\ref{pb}})\textit{%
\ is said to be \textbf{reducible }if this Poisson bracket is equivalent to
the Dubrovin-Novikov bracket }(\textbf{\ref{dn}})\textit{\ under the some
restriction }$A^{k}=f^{k}(\mathbf{u})$\textit{.}

\textbf{Conjecture 2}: Possibly, one can find appropriate constraints $%
A^{k}=f^{k}(\mathbf{u})$ for any coefficients $\mathcal{W}^{kn}(\mathbf{A})$%
, that the Poisson bracket (\textbf{\ref{pb}}) becomes \textit{reducible} to
the Dubrovin-Novikov bracket (\textbf{\ref{dn}}) in infinitely many ways.

\textbf{Example 1}: The Kupershmidt--Manin bracket (\textbf{\ref{aca}}) is
reducible, because at least one moment decomposition $A^{k}=\Sigma
(a^{m})^{k}b^{m}$ exists (see (\textbf{\ref{rac}}), (\textbf{\ref{lg}})).
Below $N-1$ parametric family of a moment decomposition is described.

As usual we call the ``hydrodynamic type system'' the 1+1 quasilinear system
of nonlinear PDE's of the first order (see (\textbf{\ref{first}}) and
details in \textbf{\cite{Dubr+Nov}}). \textit{If }$N\rightarrow \infty $%
\textit{, but the number of components }$\upsilon _{j}^{i}(\mathbf{u})$%
\textit{\ is finite for every index }$i$\textit{\ and each of these
components is a function of finite number of field variables }$u^{k}$\textit{%
, then such hydrodynamic type systems we call ``hydrodynamic chains''} (see
details in \textbf{\cite{Fer+Dav}}, \textbf{\cite{Maks+Egor}}, \textbf{\cite%
{Maks+eps}}). \textit{Also, we re-numerate equations of hydrodynamic chains
in order that every row determined by the index }$i$\textit{\ depends on a
number of field variables larger than }$i$. Thus, the hydrodynamic chain is
an infinite component generalization of a hydrodynamic type system%
\begin{equation}
A_{t}^{k}=\underset{n=0}{\overset{N_{k}}{\sum }}V_{n}^{k}(\mathbf{A}%
)A_{x}^{n}\text{, \ \ \ \ \ \ }k=1,2,...,  \label{cha}
\end{equation}%
where $V_{n}^{k}(\mathbf{A})$ are functions of the moments $A^{m}$, $%
m=1,2,...M_{k}$; $M_{k}$ and $N_{k}$ are some integers.

Let us introduce the infinite set of the moments $A^{k}=f^{k}(\mathbf{u})$,
where $f^{k}(\mathbf{u})$ are some functions. For simplicity, below we shall
describe hydrodynamic type systems (\textbf{\ref{first}}) whose \textit{%
coefficients }$\upsilon _{j}^{i}(\mathbf{u})$\textit{\ depend explicitly on
number }$N$\textit{\ of field variables }$u^{k}$.

\textbf{Example 2}: The Zakharov reduction (\textbf{\ref{sw}}) explicitly
depends on number of the field variables $b^{k}$ (because $A^{0}=\Sigma
b^{k} $), but the Benney hydrodynamic chain does not depend on this number.

If the hydrodynamic type system (\textbf{\ref{first}}) can be written via
these moments $A^{k}$, whose \textit{coefficients }$V_{j}^{i}(\mathbf{A})$%
\textit{\ are \textbf{independent} on the above number }$N$\textit{, then
such a hydrodynamic type system }(\textbf{\ref{first}})\textit{\ we call a 
\textbf{hydrodynamic reduction} (see definition and details below) of the 
\textbf{hydrodynamic chain }}(\textbf{\ref{cha}}). The variables $a^{k}$ are
said to be ``flat'' coordinates, the hydrodynamic type system (\textbf{\ref%
{first}}) is said to be written in the \textit{canonical} form (\textbf{\ref%
{loc}}), if the Christoffel symbols are vanished (see \textbf{\cite{Dubr+Nov}%
}, \textbf{\cite{Tsar}}).

Let us re-compute the canonical Poisson bracket (see (\textbf{\ref{loc}}))%
\begin{equation}
\{a^{i},a^{j}\}=\bar{g}^{ij}\delta ^{\prime }(x-x^{\prime })  \label{can}
\end{equation}%
via moments $A^{k}$:%
\begin{equation}
\{A^{k},A^{n}\}=\frac{\partial f^{k}(\mathbf{a})}{\partial a^{s}}\bar{g}%
^{sm}\partial _{x}\frac{\partial f^{n}(\mathbf{a})}{\partial a^{m}}\delta
(x-x^{\prime }).  \label{rec}
\end{equation}%
Assume the above r.h.s. can be expressed via moments $A^{k}$ only, i.e. the
coefficients $\mathcal{W}^{kn}(\mathbf{A})$ in (\textbf{\ref{pb}}) are 
\textbf{independent} on number $N$ (see the examples below).

\textbf{Definition 2}: \textit{The hydrodynamic chain }(\textbf{\ref{cha}})%
\textit{\ written in the form} (\textbf{\ref{pb}})%
\begin{equation}
A_{t}^{k}=\{A^{k},\mathbf{\bar{H}}\}=[\mathcal{W}^{kn}(\mathbf{A})\partial
_{x}+\partial _{x}\mathcal{W}^{nk}(\mathbf{A})]\frac{\delta \mathbf{\bar{H}}%
}{\delta A^{n}}\text{, \ \ \ \ \ }k,n=0,1,2,...  \label{!}
\end{equation}%
\textit{\ is said to be Hamiltonian.}

\textbf{Remark}: The above local Hamiltonian structures (\textbf{\ref{!}})\
and corresponding Poisson brackets still are not investigated properly.
However, plenty publications are devoted to their particular cases (see, for
instance, \textbf{\cite{Dorfman}}, \textbf{\cite{Kuper}}, \textbf{\cite{KM}}%
), some of them will be described below. The classification of so-called $M$
brackets is given in \textbf{\cite{Maks+Puas}}. All examples presented in
this paper belong to this class.

Local Hamiltonian structures (\textbf{\ref{!}}) for hydrodynamic chains we
call \textit{reducible} if they are obtained by the above procedure directly
from local Hamiltonian structures (see (\textbf{\ref{can}})) written in flat
coordinates of corresponding hydrodynamic reductions. So, in this paper we
describe a broad class of hydrodynamic chains associated with reducible
Poisson brackets. All other Poisson brackets (\textbf{\ref{!}}) and
corresponding integrable hydrodynamic chains will be discussed elsewhere.

Let us, for instance, introduce the infinite set of moments $A^{k}$ via
multi-index constant matrices $\mathbf{\varepsilon }$%
\begin{equation*}
A^{1}=\varepsilon _{k}^{(0)}a^{k}+\varepsilon ^{0}\text{, \ \ \ }%
A^{2}=\varepsilon _{jk}^{(0)}a^{j}a^{k}+\varepsilon
_{k}^{(1)}a^{k}+\varepsilon ^{1}\text{, \ \ \ }A^{3}=\varepsilon
_{ijk}^{(0)}a^{i}a^{j}a^{k}+\varepsilon _{jk}^{(1)}a^{j}a^{k}+\varepsilon
_{k}^{(2)}a^{k}+\varepsilon ^{2},...,
\end{equation*}%
where, obviously, all multi-index matrices $\mathbf{\varepsilon }$ are
symmetric under a permutation of their sub-indexes. Then%
\begin{eqnarray*}
\{A^{1},A^{1}\} &=&\frac{\partial A^{0}}{\partial a^{i}}\bar{g}^{ij}\partial
_{x}\frac{\partial A^{0}}{\partial a^{j}}\delta (x-x^{\prime })=\varepsilon
_{k}^{(0)}\bar{g}^{km}\varepsilon _{m}^{(0)}\delta ^{\prime }(x-x^{\prime })%
\text{,} \\
&& \\
\{A^{2},A^{1}\} &=&\frac{\partial A^{1}}{\partial a^{i}}\bar{g}^{ij}\partial
_{x}\frac{\partial A^{0}}{\partial a^{j}}\delta (x-x^{\prime
})=[2\varepsilon _{m}^{(0)}\bar{g}^{ms}\varepsilon
_{sk}^{(0)}a^{k}+\varepsilon _{m}^{(1)}\bar{g}^{ms}\varepsilon
_{s}^{(0)}]\delta ^{\prime }(x-x^{\prime })\text{,} \\
&& \\
\{A^{2},A^{2}\} &=&\frac{\partial A^{1}}{\partial a^{i}}\bar{g}^{ij}\partial
_{x}\frac{\partial A^{1}}{\partial a^{j}}\delta (x-x^{\prime
})=[4\varepsilon _{ns}^{(0)}\bar{g}^{sm}\varepsilon
_{mk}^{(0)}a^{n}a^{k}+4\varepsilon _{m}^{(1)}\bar{g}^{ms}\varepsilon
_{sk}^{(0)}a^{k} \\
&& \\
&&+\varepsilon _{m}^{(1)}\bar{g}^{ms}\varepsilon _{s}^{(1)}]\delta ^{\prime
}(x-x^{\prime })+2[\varepsilon _{ns}^{(0)}\bar{g}^{sm}\varepsilon
_{mk}^{(0)}a^{n}a^{k}+\varepsilon _{m}^{(1)}\bar{g}^{ms}\varepsilon
_{sk}^{(0)}a^{k}]_{x}\delta (x-x^{\prime }),...
\end{eqnarray*}%
\textit{The main problem in the classification of such local Hamiltonian
structures for hydrodynamic chains is a description of all multi-index
matrices }$\mathbf{\varepsilon }$ \textit{allowing to \textbf{express
explicitly} the polynomials with respect to flat coordinates }$a^{k}$\textit{%
\ in right hand sides of the above Poisson bracket via \textbf{polynomials}
of moments }$A^{n}$. Then such moments are analogues of the Liouville
coordinates for an infinite dimensional space (see \textbf{\cite{Dubr+Nov}}, 
\textbf{\cite{Malt+Nov}},\textbf{\ \cite{Maks+Puas}}), i.e.%
\begin{eqnarray*}
\{A^{0},A^{0}\} &=&\alpha ^{00}\delta ^{\prime }(x-x^{\prime })\text{, \ \ \
\ \ }\{A^{1},A^{0}\}=[\alpha ^{10}A^{0}+\alpha ^{11}]\delta ^{\prime
}(x-x^{\prime }), \\
&& \\
\{A^{1},A^{1}\} &=&[(\alpha ^{20}A^{1}+\alpha ^{21}(A^{0})^{2}+\alpha
^{22})\partial _{x}+\partial _{x}(\alpha ^{20}A^{1}+\alpha
^{21}(A^{0})^{2}+\alpha ^{22})]\delta (x-x^{\prime }),...
\end{eqnarray*}%
\textit{All coefficients }$\alpha ^{ik}$\textit{\ must be \textbf{independent%
} on} $N$.

\textbf{Example 3}: Let us introduce the moments%
\begin{equation}
A^{k}=\frac{1}{\beta k+\gamma }\underset{i=1}{\overset{N}{\sum }}\varepsilon
_{i}(a^{i})^{\beta k+\gamma }\text{, \ \ \ \ \ \ }k=0,\pm 1,\pm 2,...,
\label{simple}
\end{equation}%
where $\varepsilon _{i}$, $\beta $ and $\gamma $ are arbitrary constants.
Any symmetric non-degenerate matrix $\bar{g}^{ij}$ can be written as a
diagonal matrix under a linear change of the field variables $a^{i}$. Thus,
without lost of generality we assume that diagonal metric coefficients are $%
\alpha _{i}$ (off-diagonal coefficients are absent, see (\textbf{\ref{tot}}%
)). Then we have (see (\textbf{\ref{rec}}))%
\begin{equation*}
\{A^{k},A^{n}\}=\underset{i=1}{\overset{N}{\sum }}\alpha _{i}\varepsilon
_{i}^{2}[(a^{i})^{\beta (k+n)+2(\gamma -1)}\partial _{x}+(\beta n+\gamma
-1)(a^{i})^{\beta (k+n)+2(\gamma -1)}a_{x}^{i}]\delta (x-x^{\prime }).
\end{equation*}%
In general case r.h.s. can be expressed via moments $A^{k}$ in \textbf{%
infinitely many ways} for every \textit{fixed} number $N$. However, in the
above construction r.h.s. cannot depend on number $N$. Since $A^{k}$ is a
sum of monomials of the sole degree $\beta k+\gamma $, then the sum of
monomials of another sole degree $\beta (k+n)+2(\gamma -1)$ must be
expressed via the moments $A^{k+n+m}$, where $m$ is an integer, which is not
determined yet. Then $\alpha _{i}=1/\varepsilon _{i}$ and%
\begin{equation*}
\{A^{k},A^{n}\}=[(\beta (k+m)+1)A^{k+n+m}\partial _{x}+(\beta
(n+m)+1)\partial _{x}A^{k+n+m}]\delta (x-x^{\prime }),
\end{equation*}%
where $\gamma =\beta m+2$. Since $\beta \neq 0$, one can re-scale all
moments $A^{k}\rightarrow \beta A^{k}$. Introducing new parameter $l=1/\beta 
$, one can obtain the two parametric\ family of the Poisson brackets%
\begin{equation*}
\{A_{(m,l)}^{k},A_{(m,l)}^{n}\}=[(k+m+l)A_{(m,l)}^{k+n+m}\partial
_{x}+(n+m+l)\partial _{x}A_{(m,l)}^{k+n+m}]\delta (x-x^{\prime }),
\end{equation*}%
where%
\begin{equation}
A_{(m,l)}^{k}=\frac{l^{2}}{k+m+2l}\underset{i=1}{\overset{N}{\sum }}%
\varepsilon _{i}(a^{i})^{\frac{k+m}{l}+2}\text{, \ \ \ \ \ \ }k,m\in \mathbf{%
Z}  \label{bla}
\end{equation}

\textbf{Remark}: If also $l\in \mathbf{Z}$, then the some value of running
index $k^{\ast }=-(m+2l)$ determines the corresponding moment%
\begin{equation*}
A_{(m,l)}^{-m-2l}=l\underset{i=1}{\overset{N}{\sum }}\varepsilon _{i}\ln
a^{i}.
\end{equation*}

\textbf{Remark}: The above Poisson bracket can be obtained with the aid of
another moment decomposition (cf. (\textbf{\ref{rac}}))%
\begin{equation*}
A_{(m,l)}^{k}=l\underset{i=1}{\overset{N}{\sum }}(a^{i})^{\frac{k+m}{l}%
+1}b^{i},
\end{equation*}%
where another canonical Poisson bracket (\textbf{\ref{lg}}).

\textbf{Example 4}: Let us consider the \textit{nonlocal} Poisson bracket
(see \textbf{\cite{Fer+Mokh}})%
\begin{equation*}
\{a^{i},a^{j}\}=\left[ \left( \frac{\delta _{ij}}{\varepsilon _{i}}%
-\varepsilon a^{i}a^{j}\right) \partial _{x}-\varepsilon
a^{j}a_{x}^{i}+\varepsilon a_{x}^{i}\partial _{x}^{-1}a_{x}^{j}\right]
\delta (x-x^{\prime }),
\end{equation*}%
where $\delta _{ij}$ is the Kronecker symbol and $\varepsilon $ is a
constant curvature. Introducing the moments $A^{k}$ with the aid of (\textbf{%
\ref{simple}}), the above Poisson bracket leads to the Poisson bracket%
\begin{equation*}
\{A^{k},A^{n}\}=[(\beta (k+m)+1)A^{k+n+m}\partial _{x}+(\beta
(n+m)+1)\partial _{x}A^{k+n+m}]\delta (x-x^{\prime })
\end{equation*}%
\begin{equation*}
-\varepsilon \lbrack (\beta (k+m)+2)(\beta (n+m)+2)A^{k}A^{n}\partial
_{x}+(\beta (k+m)+2)(\beta (n+m)+1)A^{k}A_{x}^{n}
\end{equation*}%
\begin{eqnarray*}
&& \\
&&+(\beta (n+m)+2)A^{n}A_{x}^{k}-A_{x}^{k}\partial _{x}^{-1}A_{x}^{n}]\delta
(x-x^{\prime }),
\end{eqnarray*}%
where $\gamma =\beta m+2$.

\textbf{Example 5}: Let us introduce the moments (cf. (\textbf{\ref{simple}}%
))%
\begin{equation*}
A^{k}=\frac{1}{\beta k+\gamma }\underset{i=1}{\overset{N}{\sum }}\varepsilon
_{i}[f(a^{i})]^{\beta k+\gamma }\text{, \ \ \ \ \ \ }k=0,\pm 1,\pm 2,...,
\end{equation*}%
where%
\begin{equation*}
f^{\prime ^{2}}(z)=R_{M}(f)\equiv \alpha _{0}f^{M}+\alpha _{1}f^{M-1}+\alpha
_{2}f^{M-2}+...+\alpha _{M}.
\end{equation*}%
Then we have%
\begin{equation*}
\{A^{k},A^{n}\}=\underset{i=1}{\overset{N}{\sum }}\varepsilon
_{i}[f(a^{i})]^{\beta k+\gamma -1}f^{\prime }(a^{i})\partial
_{x}[f(a^{i})]^{\beta n+\gamma -1}f^{\prime }(a^{i})\delta (x-x^{\prime }).
\end{equation*}%
Expressing r.h.s. for some special set of parameters $\beta $ and $\gamma $
via higher moments, one can obtain exactly\ the Dorfman Poisson brackets
investigated in \textbf{\cite{Dorfman}}.

If, for instance (cf. (\textbf{\ref{bla}})),%
\begin{equation*}
A_{(m,n)}^{n-2m}=(m-n)\underset{i=1}{\overset{N}{\sum }}\varepsilon _{i}\ln
\wp (a^{i})\text{, \ \ \ }A_{(m,n)}^{k}|_{k\neq n-2m}=\frac{(m-n)^{2}}{k+2m-n%
}\underset{i=1}{\overset{N}{\sum }}\varepsilon _{i}\wp ^{\frac{k+2m-n}{m-n}%
}(a^{i})\text{, \ \ \ }m,n\in \mathbf{Z}.
\end{equation*}%
where $\wp (z,g_{2},g_{3})$ is the Weiershtrass elliptic function, then
corresponding Poisson bracket (\textbf{\ref{pb}}) is determined by%
\begin{equation*}
B^{ks}(\mathbf{A})=(4k+10m-6n)A_{(m,n)}^{k+s+3m-2n}-(k+\frac{3m-n}{2}%
)g_{2}A_{(m,n)}^{k+s+m}-(k+m)g_{3}A_{(m,n)}^{k+s+n}.
\end{equation*}

\textbf{Example 6}: Let us introduce the moments%
\begin{equation*}
A^{k}=\underset{i=1}{\overset{N}{\sum }}\varepsilon _{i}\int
U^{k}(a^{i})da^{i},
\end{equation*}%
where%
\begin{equation}
U^{\prime }=\underset{m=0}{\overset{M}{\sum }}\gamma _{m}U^{m}.  \label{e}
\end{equation}%
Then we have%
\begin{eqnarray*}
\{A^{k},A^{n}\} &=&\underset{i=1}{\overset{N}{\sum }}\varepsilon
_{i}U^{k}(a^{i})\partial _{x}U^{n}(a^{i})\delta (x-x^{\prime }) \\
&& \\
&=&\underset{i=1}{\overset{N}{\sum }}\varepsilon _{i}U^{k+n}(a^{i})\delta
^{\prime }(x-x^{\prime })+\frac{n}{k+n}\left[ \underset{i=1}{\overset{N}{%
\sum }}\varepsilon _{i}U^{k+n}(a^{i})\right] _{x}\delta (x-x^{\prime }).
\end{eqnarray*}%
Since (see (\textbf{\ref{e}}))%
\begin{equation*}
\frac{1}{k+1}\underset{i=1}{\overset{N}{\sum }}\varepsilon
_{i}U^{k+1}(a^{i})=\underset{m=0}{\overset{M}{\sum }}\gamma _{m}A^{m+k},
\end{equation*}%
then this Poisson bracket can be written in the most compact form%
\begin{equation*}
\{A^{0},A^{0}\}=\underset{n=1}{\overset{N}{\sum }}\varepsilon _{n}\delta
^{\prime }(x-x^{\prime })\text{, \ \ \ \ \ \ }\{A^{k},A^{n}\}=\underset{m=0}{%
\overset{M}{\sum }}\gamma _{m}[kA^{m+k+n-1}\partial _{x}+n\partial
_{x}A^{m+k+n-1}]\delta (x-x^{\prime }).
\end{equation*}%
These are exactly the Dorfman Poisson brackets (see \textbf{\cite{Dorfman}}).

\textbf{Remark}: all above Poisson brackets are obtained by the
phenomenological Hamiltonian approach presented in this paper. However, one
can verify directly (see \textbf{\cite{Maks+Puas}}), that, indeed, these
Poisson brackets satisfy the Jacobi identity (\textbf{\ref{jac}}).

\section{The integrability criteria}

In this section we establish a new approach in the classification of
integrable hydrodynamic chains and their integrability. If a hydrodynamic
type system possesses the Hamiltonian structure, it is not enough for
integrability (the ``integrability'' means the existence of infinitely many
conservation laws and commuting flows (see \textbf{\cite{Tsar}}). However,
integrable hydrodynamic chains possessing the Hamiltonian structure can be
classified. The Hamiltonian hydrodynamic reductions determine a generating
function of conservation laws, which exists for integrable hydrodynamic
chains only.

\textbf{Theorem 2}: \textit{If a hydrodynamic chain has a reducible Poisson
bracket, then this Hamiltonian hydrodynamic chain has at least one }$N-1$ 
\textit{parametric family of the Hamiltonian hydrodynamic reductions}.

\textbf{Proof}: Without lost of generality we restrict our consideration on
the Kupershmidt--Manin bracket (\textbf{\ref{aca}}). Assume for simplicity,
that the Hamiltonian is $\mathbf{\bar{H}}_{2}=\int \mathbf{H}%
_{2}(A^{0},A^{1},A^{2})dx$. Then the corresponding hydrodynamic chain (%
\textbf{\ref{bhs}}) is%
\begin{equation}
A_{t}^{k}=(kA^{k+1}\partial _{x}+2\partial _{x}A^{k+1})\frac{\partial 
\mathbf{H}_{2}}{\partial A^{2}}+(kA^{k}\partial _{x}+\partial _{x}A^{k})%
\frac{\partial \mathbf{H}_{2}}{\partial A^{1}}+kA^{k-1}\partial _{x}\frac{%
\partial \mathbf{H}_{2}}{\partial A^{0}}.  \label{b}
\end{equation}%
This hydrodynamic chain has the moment decomposition (\textbf{\ref{simple}}%
), where $\beta =\gamma =1$. Then the above hydrodynamic chain reduces to
the hydrodynamic type system written in the conservative form%
\begin{equation}
a_{t}^{i}=\partial _{x}\left( \frac{\partial \mathbf{H}_{2}}{\partial A^{0}}%
+a^{i}\frac{\partial \mathbf{H}_{2}}{\partial A^{1}}+(a^{i})^{2}\frac{%
\partial \mathbf{H}_{2}}{\partial A^{2}}\right) .  \label{sis}
\end{equation}%
This hydrodynamic chain has at least three local conservation laws%
\begin{eqnarray*}
A_{t}^{0} &=&\partial _{x}\left( A^{0}\frac{\partial \mathbf{H}_{2}}{%
\partial A^{1}}+2A^{1}\frac{\partial \mathbf{H}_{2}}{\partial A^{2}}\right) 
\text{, \ \ \ \ }A_{t}^{1}=\partial _{x}\left( A^{0}\frac{\partial \mathbf{H}%
_{2}}{\partial A^{0}}+2A^{1}\frac{\partial \mathbf{H}_{2}}{\partial A^{1}}%
+3A^{2}\frac{\partial \mathbf{H}_{2}}{\partial A^{2}}-\mathbf{H}_{2}\right) ,
\\
&& \\
\partial _{t}\mathbf{H}_{2} &=&\partial _{x}\left( 2A^{3}(\frac{\partial 
\mathbf{H}_{2}}{\partial A^{2}})^{2}+3A^{2}\frac{\partial \mathbf{H}_{2}}{%
\partial A^{1}}\frac{\partial \mathbf{H}_{2}}{\partial A^{2}}+A^{1}(2\frac{%
\partial \mathbf{H}_{2}}{\partial A^{0}}\frac{\partial \mathbf{H}_{2}}{%
\partial A^{2}}+(\frac{\partial \mathbf{H}_{2}}{\partial A^{1}})^{2})+A^{0}%
\frac{\partial \mathbf{H}_{2}}{\partial A^{0}}\frac{\partial \mathbf{H}_{2}}{%
\partial A^{1}}\right) .
\end{eqnarray*}%
Then the hydrodynamic reduction (\textbf{\ref{sis}}) must have the same
conservation laws. From the first of them one can obtain the following
restriction $\Sigma \varepsilon _{m}=0$. Thus, $N-1$ parametric family of
hydrodynamic reductions is found. The above hydrodynamic reduction has the
Hamiltonian form (\textbf{\ref{ham}}) while the Hamiltonian density of
hydrodynamic type system $\mathbf{h}_{2}$ is the \textit{reduced}
Hamiltonian density $\mathbf{H}_{2}$ of hydrodynamic chain (\textbf{\ref{b}}%
), where $A^{0}=\Sigma \varepsilon _{m}a^{m}$, $A^{1}=\Sigma \varepsilon
_{m}(a^{m})^{2}/2$, $A^{2}=\Sigma \varepsilon _{m}(a^{m})^{3}/3$. Then one
can check the identity%
\begin{eqnarray*}
d\mathbf{h}_{2} &=&\underset{i=1}{\overset{N}{\sum }}\frac{\partial \mathbf{h%
}_{2}}{\partial a^{i}}da^{i}=\underset{i=1}{\overset{N}{\sum }}\varepsilon
_{i}\left( \frac{\partial \mathbf{H}_{2}}{\partial A^{0}}+a^{i}\frac{%
\partial \mathbf{H}_{2}}{\partial A^{1}}+(a^{i})^{2}\frac{\partial \mathbf{H}%
_{2}}{\partial A^{2}}\right) da^{i} \\
&& \\
&=&\underset{i=1}{\overset{N}{\sum }}\left( \frac{\partial \mathbf{H}_{2}}{%
\partial A^{0}}dA^{0}+\frac{\partial \mathbf{H}_{2}}{\partial A^{1}}dA^{1}+%
\frac{\partial \mathbf{H}_{2}}{\partial A^{2}}dA^{2}\right) =d\mathbf{H}_{2}.
\end{eqnarray*}%
Thus, indeed, $N-1$ parametric family of hydrodynamic reductions has the
Hamiltonian form, and $a^{k}$ are flat coordinates.

The \textbf{main statement}: \textit{if the hydrodynamic chain} (\textbf{\ref%
{b}}) \textit{is integrable, then above hydrodynamic reductions} (\textbf{%
\ref{sis}}) \textit{determine the generating function of conservation laws}%
\begin{equation}
p_{t}=\partial _{x}\left( \frac{\partial \mathbf{H}_{2}}{\partial A^{2}}%
p^{2}+\frac{\partial \mathbf{H}_{2}}{\partial A^{1}}p+\frac{\partial \mathbf{%
H}_{2}}{\partial A^{0}}\right)  \label{genn}
\end{equation}%
\textit{by the replacement} $a^{i}\rightarrow p$. \textit{A deformation of
the Riemann surface determined by the equation} $\lambda =\lambda (\mathbf{A}%
;p)$ \textit{is described by the Gibbons equation}%
\begin{equation*}
\lambda _{t}-\left( 2\frac{\partial \mathbf{H}_{2}}{\partial A^{2}}p+\frac{%
\partial \mathbf{H}_{2}}{\partial A^{1}}\right) \lambda _{x}=\frac{\partial
\lambda }{\partial p}\left[ p_{t}-\partial _{x}\left( \frac{\partial \mathbf{%
H}_{2}}{\partial A^{2}}p^{2}+\frac{\partial \mathbf{H}_{2}}{\partial A^{1}}p+%
\frac{\partial \mathbf{H}_{2}}{\partial A^{0}}\right) \right] .
\end{equation*}

Thus, the \textit{problem of the classification of integrable Hamiltonian
hydrodynamic chains is reduced to the problem of the classification of
generating functions of conservation laws} (see the section ``General case''
in \textbf{\cite{Maks+vech}}).

Here we enumerate \textbf{four} different tools allowing to prove an
integrability of such hydrodynamic chains.

\textbf{1}. \textit{Extra conservation law and method of pseudopotentials}.
For instance, suppose the hydrodynamic chain (\textbf{\ref{b}}) is
integrable, then the first commuting flow determined by the Hamiltonian
density $\mathbf{H}_{3}(A^{0},A^{1},A^{2},A^{3})$ creates the corresponding
generating function of conservation laws is%
\begin{equation}
p_{t^{2}}=\partial _{x}\left( \frac{\partial \mathbf{H}_{3}}{\partial A^{3}}%
p^{3}+\frac{\partial \mathbf{H}_{3}}{\partial A^{2}}p^{2}+\frac{\partial 
\mathbf{H}_{3}}{\partial A^{1}}p+\frac{\partial \mathbf{H}_{3}}{\partial
A^{0}}\right) .  \label{gem}
\end{equation}%
The compatibility condition $\partial _{t^{2}}(p_{t})=\partial
_{t}(p_{t^{2}})$ leads to exactly the same system of nonlinear PDE's in
involution as obtained by the tensor approach in \textbf{\cite{Fer+Dav}}.
The equations (\textbf{\ref{genn}}) and (\textbf{\ref{gem}}) are nothing but
pseudopotentials for corresponding 2+1 quasilinear equation (see, for
instance, (\textbf{\ref{gen}}) -- the compatibility condition $\partial
_{t^{2}}(p_{t})=\partial _{t}(p_{t^{2}})$ yields the Khohlov--Zabolotzkaya
equation, see (\textbf{\ref{bm}}) and below). Thus, we \textit{extend} the
concept of pseudopotentials on the theory of integrable hydrodynamic chains
(see the section \textbf{5}).

\textbf{2}. \textit{Reciprocal transformations}. The Hamiltonian density $%
\mathbf{H}_{2}$ can be \textit{linear} $\mathbf{H}_{2}=A^{2}+f(A^{0},A^{1})$%
, \textit{quasilinear} $\mathbf{H}_{2}=g(A^{0},A^{1})A^{2}+f(A^{0},A^{1})$
and \textit{fully nonlinear} $\mathbf{H}_{2}(A^{0},A^{1},A^{2})$. Thus, the
generating functions (\textbf{\ref{genn}}) can be separated on three
sub-classes%
\begin{eqnarray*}
p_{t} &=&\partial _{x}\left( p^{2}+\frac{\partial f(A^{0},A^{1})}{\partial
A^{1}}p+\frac{\partial f(A^{0},A^{1})}{\partial A^{0}}\right) , \\
&& \\
p_{t} &=&\partial _{x}\left[ g(A^{0},A^{1})p^{2}+\left( \frac{\partial
g(A^{0},A^{1})}{\partial A^{1}}p+\frac{\partial g(A^{0},A^{1})}{\partial
A^{0}}\right) A^{2}+\frac{\partial f(A^{0},A^{1})}{\partial A^{1}}p+\frac{%
\partial f(A^{0},A^{1})}{\partial A^{0}}\right] , \\
&& \\
p_{t} &=&\partial _{x}\left( \frac{\partial \mathbf{H}_{2}}{\partial A^{2}}%
p^{2}+\frac{\partial \mathbf{H}_{2}}{\partial A^{1}}p+\frac{\partial \mathbf{%
H}_{2}}{\partial A^{0}}\right) .
\end{eqnarray*}%
The difference between these sub-classes is obvious. The \textbf{main claim}
is that the factors $g(A^{0},A^{1})$ and $\partial \mathbf{H}_{2}/\partial
A^{2}$ can be removed under some special reciprocal transformations. Thus,
an integrability of an arbitrary (\textit{quasilinear} or \textit{fully
nonlinear}) Hamiltonian integrable hydrodynamic chain is reduced to an
integrability of a \textit{linear }Hamiltonian integrable hydrodynamic chain
(see the sections \textbf{6} and \textbf{7}).

\textbf{3}. \textit{Asymptotic at the vicinity of each singular point}.
Suppose we have some generating function of conservation laws of polynomial
type (see, for instance, the above \textit{linear} case)%
\begin{equation*}
p_{t}=\partial _{x}\left( \frac{p^{N}}{N}%
+a_{0}p^{N-1}+a_{1}p^{N-2}+...+a_{N-1}\right) .
\end{equation*}%
Then the generation function of conservation law densities $p$ can be
determined by the series%
\begin{equation}
p=\lambda +\mathbf{H}_{0}+\frac{\mathbf{H}_{1}}{\lambda }+\frac{\mathbf{H}%
_{2}}{\lambda ^{2}}+\frac{\mathbf{H}_{3}}{\lambda ^{3}}+...  \label{series}
\end{equation}%
If a hydrodynamic chain is integrable, then such an expansion exists (see
the section \textbf{7}).

\textbf{4}. \textit{Hydrodynamic reductions}. This is the most universal
approach (see \textbf{\cite{Fer+Kar}}, \textbf{\cite{Gib+Tsar}}, \textbf{%
\cite{Maks+vech}}). Let us consider the equation of the Riemann surface $%
\lambda =\lambda (\mathbf{A};p)$ satisfying the Gibbons equation (cf. (%
\textbf{\ref{gib}}); see \textbf{\cite{Maks+vech}})%
\begin{equation*}
\lambda _{t}-\frac{\partial \psi }{\partial p}\lambda _{x}=\frac{\partial
\lambda }{\partial p}[p_{t}-\partial _{x}\psi (\mathbf{A};p)].
\end{equation*}%
If $\lambda =\limfunc{const}$, then the Gibbons equation reduces to the
generating function of conservation laws $p_{t}=\partial _{x}\psi (\mathbf{A}%
;p)$. Suppose the moments $A^{n}$ are functions of $N$ Riemann invariants $%
r^{k}$; i.e. we seek $N$ component hydrodynamic reductions written in the
diagonal form%
\begin{equation*}
r_{t}^{i}=\mu ^{i}(\mathbf{r})r_{x}^{i},
\end{equation*}%
where $\mu ^{i}=\partial \psi /\partial p|_{p=p^{i}}$, $r^{i}=\lambda
|_{p=p^{i}}$ and $N$ \textit{distinct} values $p^{k}$ are determined by the
algebraic equation $\partial \lambda /\partial p=0$. Then a consistency of
the above hydrodynamic type system and its generating function of
conservation laws yields the extended Gibbons--Tsarev system (see \textbf{%
\cite{Maks+vech}}) describing all admissible functions $\psi (\mathbf{A};p)$
(see details below). This approach is adopted for the special class of the
hydrodynamic reductions%
\begin{equation*}
a_{t}^{i}=\partial _{x}\psi (\mathbf{A};a^{i}),
\end{equation*}%
which has the Hamiltonian form (\textbf{\ref{loc}}). In this case all
moments $A^{n}$ can be explicitly expressed via flat coordinates $a^{k}$.
The method of $N$ component hydrodynamic reductions (see (\textbf{\ref{gt}})
in the Riemann invariants $r^{k}$ and (\textbf{\ref{egt}}) in the
conservation law densities $a^{i}$) leads to the Gibbons--Tsarev system,
whose general solution is parameterized by $N$ arbitrary functions of a
single variable. The method of $N$ component \textit{Hamiltonian}
hydrodynamic reductions \textbf{reduces the Gibbons--Tsarev system to the 
\textit{sole} ODE} (see for instance (\textbf{\ref{water}})). Let us
consider, for instance, (\textbf{\ref{egt}}). If the hydrodynamic reduction (%
\textbf{\ref{for}}) has the Hamiltonian structure (\textbf{\ref{ham}}), then
the moment $A^{0}$ is a some function of $\Delta =\Sigma \varepsilon
_{n}a^{n}$. Thus, the system of nonlinear PDE's (\textbf{\ref{egt}}) reduces
to the ODE $(A^{0})^{\prime \prime }=0$. Thus, indeed $A^{0}=\Sigma
\varepsilon _{n}a^{n}$ in accordance with (\textbf{\ref{water}}).

Thus, the Hamiltonian approach consists of following steps:

\textbf{1}. A ``moment decomposition'' search for given Poisson bracket (see
the previous section).

\textbf{2}. A transformation to the \textit{linear} Hamiltonian case.

\textbf{3}. An asymptotic investigation of generating functions of
conservation laws.

\textbf{4}. A derivation of the \textit{extended} Gibbons--Tsarev system in
involution describing ``integrable'' Hamiltonians and $N$ component
hydrodynamic reductions parameterized by $N$ arbitrary functions of a single
variable (see \textbf{\cite{Maks+vech}}).

\section{Classification of simplest Hamiltonian hydrodynamic chains}

The Kupershmidt Poisson brackets%
\begin{equation}
\{A^{k},A^{n}\}=[(k+\gamma )A^{k+n-M}\partial _{x}+(n+\gamma )\partial
_{x}A^{k+n-M}]\delta (x-x^{\prime }),  \label{dorf}
\end{equation}%
where $M$ is an arbitrary integer, is a particular case of so-called $M$%
--brackets (see \textbf{\cite{Maks+Puas}})%
\begin{equation*}
\{A^{k},A^{n}\}=[B^{k,n}\partial _{x}+\partial _{x}B^{n,k}]\delta
(x-x^{\prime }),
\end{equation*}%
where the coefficients $B^{k,n}$ depend on the first $k+n-M$ elements only
(if $k+n<M$, then corresponding coefficients are constants). These Poisson
brackets are equivalent to each other under the re-numeration $%
A^{k}\rightarrow A^{k+s}$ (where $s$ is an appropriate integer). The case $%
M=1$ was considered in \textbf{\cite{KM}}, the case $M=0$ was considered in 
\textbf{\cite{Kuper}}, the case $M=-1$ was considered in \textbf{\cite%
{Kuper1}}.

In this section the classification of integrable hydrodynamic chains
determined by the Kupershmidt Poisson brackets (see (\textbf{\ref{dorf}}), $%
M=-1$ and \textbf{\cite{Kuper1}}, $\gamma \neq 0$)%
\begin{equation*}
C_{t^{1}}^{k}=[(k+\gamma )C^{k+n+1}\partial _{x}+(n+\gamma )\partial
_{x}C^{k+n+1}]\frac{\delta \mathbf{\bar{H}}_{0}}{\delta C^{n}}\text{, \ \ \
\ \ }k=0,1,2,...,
\end{equation*}%
where $\mathbf{\bar{H}}_{0}\mathbf{=}\int \mathbf{H}_{0}(C^{0})dx$, is
given. If the Hamiltonian density $\mathbf{H}_{0}(C^{0})$ is an \textit{%
arbitrary} function, then the hydrodynamic chain%
\begin{equation}
C_{t^{1}}^{k}=\gamma \mathbf{H}_{0}^{\prime }(C^{0})C_{x}^{k+1}+(k+2\gamma
)C^{k+1}\mathbf{H}_{0}^{\prime \prime }(C^{0})C_{x}^{0}\text{, \ \ \ \ \ }%
k=0,1,2,...  \label{ak}
\end{equation}%
has just the conservation law of the energy%
\begin{equation*}
\partial _{t^{1}}\mathbf{H}_{0}(C^{0})=\partial _{x}[\gamma \mathbf{H}%
_{0}^{\prime ^{2}}(C^{0})C^{1}].
\end{equation*}

\subsection{The first integrability criterion (extra conservation law)}

\textbf{Conjecture 3}: If a hydrodynamic chain has an \textit{extra}
conservation law, then this hydrodynamic chain is integrable, i.e. this
hydrodynamic chain has infinitely many conservation laws and commuting flows
(see examples in \textbf{\cite{Kuper}}).

Indeed, a second conservation law can be found in the form%
\begin{equation*}
\partial _{t^{1}}[f(C^{0})C^{1}]=\partial _{x}\left[ \frac{\gamma }{2}%
f^{\prime }(C^{0})\mathbf{H}_{0}^{\prime }(C^{0})(C^{1})^{2}+\gamma f(C^{0})%
\mathbf{H}_{0}^{\prime }(C^{0})C^{2}\right] ,
\end{equation*}%
where%
\begin{equation}
3f^{\prime }\mathbf{H}_{0}^{\prime \prime }=f^{\prime \prime }\mathbf{H}%
_{0}^{\prime }\text{, \ \ \ \ \ \ \ \ }(\gamma +1)f\mathbf{H}_{0}^{\prime
\prime }=\gamma f^{\prime }\mathbf{H}_{0}^{\prime }.  \label{fun}
\end{equation}%
Thus, the integrable hydrodynamic chain (see \textbf{\cite{Kuper1}})%
\begin{equation}
C_{t^{1}}^{k}=\frac{\gamma (1-\gamma )}{(1-2\gamma )^{2}}(C^{0})^{\frac{%
3\gamma -1}{1-2\gamma }}[(1-2\gamma )C^{0}C_{x}^{k+1}+(k+2\gamma
)C^{k+1}C_{x}^{0}]\text{, \ \ \ \ }k=0,1,2,...  \label{j}
\end{equation}%
is determined by the Hamiltonian density $\mathbf{H}_{0}=(C^{0})^{\frac{%
1-\gamma }{1-2\gamma }}$ in general case ($\gamma \neq -1$, $\gamma \neq 1$
and $\gamma \neq 1/2$). Then the second conservation law density $\mathbf{H}%
_{1}=C^{1}(C^{0})^{\frac{1+\gamma }{1-2\gamma }}$ determines the first
commuting flow%
\begin{equation*}
C_{t^{2}}^{k}=(1+\gamma )(C^{0})^{\frac{1+\gamma }{1-2\gamma }}C_{x}^{k+2}+%
\frac{\gamma (1+\gamma )}{1-2\gamma }C^{1}(C^{0})^{\frac{3\gamma }{1-2\gamma 
}}C_{x}^{k+1}+\frac{(1+\gamma )(k+2\gamma )}{1-2\gamma }C^{k+1}(C^{0})^{%
\frac{3\gamma }{1-2\gamma }}C_{x}^{1}
\end{equation*}%
\begin{equation*}
+\frac{1+\gamma }{1-2\gamma }(C^{0})^{\frac{5\beta -1}{1-2\beta }%
}[(k+1+2\gamma )C^{0}C^{k+2}+\frac{3\gamma (k+2\gamma )}{1-2\gamma }%
C^{1}C^{k+1}]C_{x}^{0}\text{, }\ \ \ \ \ k=0,1,2,...
\end{equation*}

\textbf{Remark}: These hydrodynamic chains (for the \textbf{integer} values $%
M=(1-3\gamma )/(1-2\gamma )$) were derived in \textbf{\cite{Blaszak}} (see
also \textbf{\cite{Manasa}}) via the Lax formulation. They are first members
of so-called dispersionless limit of the $M-$dDym hierarchy. In this paper
we recover the Hamiltonian structure for these hydrodynamic chains. We
believe that the Lax formulation is connected with this Hamiltonian
structure. However, this is open question.

In particular cases:

if $\gamma =1$, then ($\mathbf{H}_{0}=\ln C^{0}$, $\mathbf{H}%
_{1}=C^{1}(C^{0})^{-2}$)%
\begin{equation}
C_{t^{1}}^{k}=\frac{1}{C^{0}}C_{x}^{k+1}-(k+2)C^{k+1}\frac{C_{x}^{0}}{%
(C^{0})^{2}}\text{, \ \ \ \ \ }k=0,1,2,...,  \label{4}
\end{equation}%
\begin{eqnarray*}
C_{t^{2}}^{k}
&=&2(C^{0})^{-2}C_{x}^{k+2}-2C^{1}(C^{0})^{-3}C_{x}^{k+1}-2(k+2)C^{k+1}(C^{0})^{-3}C_{x}^{1}
\\
&&
\end{eqnarray*}%
\begin{equation*}
-2(C^{0})^{-4}[(k+3)C^{0}C^{k+2}-3(k+2)C^{1}C^{k+1}]C_{x}^{0}\text{, \ \ \ \
\ }k=0,1,2,...;
\end{equation*}

if $\gamma =1/2$, then ($\mathbf{H}_{0}=e^{C^{0}}$, $\mathbf{H}%
_{1}=C^{1}e^{3C^{0}}$)%
\begin{equation}
C_{t^{1}}^{k}=\frac{1}{2}e^{C^{0}}C_{x}^{k+1}+(k+1)C^{k+1}e^{C^{0}}C_{x}^{0}%
\text{, \ \ \ \ \ }k=0,1,2,...,  \label{5}
\end{equation}%
\begin{eqnarray*}
C_{t^{2}}^{k} &=&\frac{3}{2}e^{3C^{0}}C_{x}^{k+2}+\frac{3}{2}%
C^{1}e^{3C^{0}}C_{x}^{k+1}+3(k+1)C^{k+1}e^{3C^{0}}C_{x}^{1} \\
&&
\end{eqnarray*}%
\begin{equation*}
+3e^{3C^{0}}[(k+2)C^{k+2}+3(k+1)C^{k+1}C^{1}]C_{x}^{0}\text{, \ \ \ \ \ }%
k=0,1,2,...;
\end{equation*}

if $\gamma =-1$, then the second conservation law%
\begin{equation*}
C_{t^{1}}^{1}=-\partial _{x}[C^{2}\mathbf{H}_{0}^{\prime }(C^{0})]\text{.}
\end{equation*}%
exists for \textbf{any} Hamiltonian density $\mathbf{H}_{0}(C^{0})$. Thus,
in this case the hydrodynamic chain%
\begin{equation*}
C_{t^{1}}^{k}=-\mathbf{H}_{0}^{\prime }(C^{0})C_{x}^{k+1}+(k-2)C^{k+1}%
\mathbf{H}_{0}^{\prime \prime }(C^{0})C_{x}^{0}\text{, \ \ \ \ \ }k=0,1,2,...
\end{equation*}%
is integrable, if the \textbf{third} conservation law%
\begin{equation*}
\partial _{t^{1}}\mathbf{H}_{2}(C^{0},C^{1},C^{2})=\partial _{x}\mathbf{G}%
_{2}(C^{0},C^{1},C^{2},C^{3})
\end{equation*}%
exists. Then,%
\begin{eqnarray*}
\mathbf{H}_{2} &=&-\frac{3}{2}(C^{0})^{1/3}C^{2}-\frac{1}{4}%
(C^{0})^{-2/3}(C^{1})^{2}, \\
&& \\
\mathbf{G}_{2} &=&C^{3}+\frac{C^{1}C^{2}}{3C^{0}}+\frac{(C^{1})^{3}}{%
27(C^{0})^{2}}
\end{eqnarray*}%
and the integrable hydrodynamic chain is%
\begin{equation*}
C_{t^{1}}^{k}=-\frac{2}{9}%
(C^{0})^{-4/3}[3C^{0}C_{x}^{k+1}+(k-2)C^{k+1}C_{x}^{0}]\text{, \ \ \ \ \ }%
k=0,1,2,...,
\end{equation*}%
where the Hamiltonian density is $\mathbf{H}_{0}=(C^{0})^{2/3}$.

\subsection{The second integrability criterion (generating functions of
conservation laws)}

Hydrodynamic reductions of the Hamiltonian chain (\textbf{\ref{j}}) one can
seek with the aid of the moment decomposition given in the form (\textbf{\ref%
{simple}})%
\begin{equation}
C^{k}=\frac{\gamma -1}{k+2\gamma -1}\underset{i=1}{\overset{N}{\sum }}%
\varepsilon _{i}(c^{i})^{\frac{k+2\gamma -1}{\gamma -1}},  \label{reda}
\end{equation}%
if $\gamma \neq (1-K)/2$, where $K=0,1,2,...$

Then the corresponding hydrodynamic type system%
\begin{equation}
c_{t^{1}}^{i}=(\gamma -1)\partial _{x}[(c^{i})^{\frac{\gamma }{\gamma -1}}%
\mathbf{h}_{0}^{\prime }(C^{0})]  \label{1}
\end{equation}%
has the local Hamiltonian structure (\textbf{\ref{ham}})%
\begin{equation}
c_{t^{1}}^{i}=\partial _{x}[\frac{\gamma -1}{\varepsilon _{i}}\frac{\delta 
\mathbf{\bar{h}}_{0}}{\delta c^{i}}],  \label{gam}
\end{equation}%
where the Hamiltonian $\mathbf{\bar{h}}_{0}=\int \mathbf{h}_{0}(C^{0})dx$.

\textbf{Remark}: If $\gamma =(1-K)/2$, then%
\begin{equation*}
C^{k}|_{k\neq K}=\frac{1+K}{2(K-k)}\underset{i=1}{\overset{N}{\sum }}%
\varepsilon _{i}(c^{i})^{\frac{2(K-k)}{1+K}}\text{, \ \ \ \ \ \ \ }C^{K}=%
\underset{i=1}{\overset{N}{\sum }}\varepsilon _{i}\ln c^{i}\text{.}
\end{equation*}%
For instance, if $\gamma =1/2$ (see (\textbf{\ref{5}})), then%
\begin{equation*}
C^{0}=\underset{i=1}{\overset{N}{\sum }}\varepsilon _{i}\ln c^{i}\text{, \ \
\ \ \ \ \ }C^{k}=-\frac{1}{2k}\underset{i=1}{\overset{N}{\sum }}\varepsilon
_{i}(c^{i})^{-2k}\text{, \ \ \ \ }k=1,2,...
\end{equation*}%
In such cases the compatibility of the above moment decomposition (\textbf{%
\ref{reda}}) with the hydrodynamic chains (\textbf{\ref{j}}) and (\textbf{%
\ref{5}}) brings to the extra parametric restriction (cf. the previous
sections)%
\begin{equation}
\underset{i=1}{\overset{N}{\sum }}\varepsilon _{i}=0.  \label{eps}
\end{equation}

If $\gamma =1$, then hydrodynamic reductions of the Hamiltonian chain (%
\textbf{\ref{4}})%
\begin{equation*}
C_{t^{1}}^{k}=\mathbf{H}_{0}^{\prime }(C^{0})C_{x}^{k+1}+(k+2)C^{k+1}\mathbf{%
H}_{0}^{\prime \prime }(C^{0})C_{x}^{0}\text{, \ \ \ \ \ }k=0\text{, }1\text{%
, }2\text{, ...}
\end{equation*}%
one can seek with the aid of the moment decomposition in the \textit{%
degenerate} form (\textbf{\ref{simple}})%
\begin{equation*}
C^{k}=\frac{1}{k+1}\underset{i=1}{\overset{N}{\sum }}\varepsilon
_{i}e^{(k+1)c^{i}}.
\end{equation*}%
Then the corresponding hydrodynamic type system%
\begin{equation}
c_{t^{1}}^{i}=\partial _{x}[\frac{e^{c^{i}}}{C^{0}}]  \label{2}
\end{equation}%
has the local Hamiltonian structure (cf. (\textbf{\ref{ham}}))%
\begin{equation*}
c_{t^{1}}^{i}=\partial _{x}[\frac{1}{\varepsilon _{i}}\frac{\delta \mathbf{%
\bar{h}}_{0}}{\delta c^{i}}],
\end{equation*}%
where the Hamiltonian $\mathbf{\bar{h}}_{0}=\int \ln C^{0}dx$.

I \textbf{postulate} the existence of the generating function of
conservation laws%
\begin{equation}
s_{t^{1}}=(\gamma -1)\partial _{x}[s^{\frac{\gamma }{\gamma -1}}\mathbf{H}%
_{0}^{\prime }(C^{0})]\text{,\ \ \ \ \ }\gamma \neq 1\text{; \ \ \ \ \ \ \ \
\ \ \ }s_{t^{1}}=\partial _{x}\left( \frac{e^{s}}{C^{0}}\right) \text{, \ \
\ }\gamma =1  \label{3}
\end{equation}%
obtained from the above hydrodynamic reductions (\textbf{\ref{1}}) and (%
\textbf{\ref{2}}) by the formal replacement $c^{i}\rightarrow s$. This is
the \textbf{main ansatz} of this Hamiltonian approach (see (\textbf{\ref%
{genn}})).

In the same way the generating function of conservation laws for the first
commuting flow can be found%
\begin{equation*}
s_{t^{2}}=(\gamma -1)\partial _{x}\left( s^{\frac{\gamma +1}{\gamma -1}}%
\frac{\partial \mathbf{H}_{1}}{\partial C^{1}}+s^{\frac{\gamma }{\gamma -1}}%
\frac{\partial \mathbf{H}_{1}}{\partial C^{0}}\right) \text{,\ \ \ \ }\gamma
\neq 1\text{; \ \ \ \ \ \ }s_{t^{2}}=\partial _{x}\left( \frac{e^{2s}}{%
(C^{0})^{2}}-2\frac{e^{s}C^{1}}{(C^{0})^{3}}\right) \text{, \ \ \ }\gamma =1.
\end{equation*}%
The substitution 2+1 quasilinear system%
\begin{eqnarray*}
C_{t^{1}}^{0} &=&2\gamma C^{1}\left( \frac{\partial \mathbf{H}_{0}}{\partial
C^{0}}\right) _{x}+\gamma \frac{\partial \mathbf{H}_{0}}{\partial C^{0}}%
C_{x}^{1}\text{, \ \ \ \ \ \ }C_{t^{1}}^{1}=(2\gamma +1)C^{2}\left( \frac{%
\partial \mathbf{H}_{0}}{\partial C^{0}}\right) _{x}+\gamma \frac{\partial 
\mathbf{H}_{0}}{\partial C^{0}}C_{x}^{2}, \\
&& \\
C_{t^{2}}^{0} &=&2\gamma C^{1}\left( \frac{\partial \mathbf{H}_{1}}{\partial
C^{0}}\right) _{x}+\gamma \frac{\partial \mathbf{H}_{1}}{\partial C^{0}}%
C_{x}^{1}+(2\gamma +1)C^{2}\left( \frac{\partial \mathbf{H}_{1}}{\partial
C^{1}}\right) _{x}+(\gamma +1)\frac{\partial \mathbf{H}_{1}}{\partial C^{1}}%
C_{x}^{2}
\end{eqnarray*}%
in the compatibility condition $(s_{t^{1}})_{t^{2}}=(s_{t^{2}})_{t^{1}}$
yields the couple of equations%
\begin{eqnarray*}
(\gamma +1)\frac{\partial \mathbf{H}_{1}}{\partial C^{1}}\frac{\partial ^{2}%
\mathbf{H}_{0}}{\partial (C^{0})^{2}} &=&\gamma \frac{\partial \mathbf{H}_{0}%
}{\partial C^{0}}\frac{\partial ^{2}\mathbf{H}_{1}}{\partial C^{0}\partial
C^{1}}\text{, \ \ \ \ \ \ \ \ \ \ }\frac{\partial ^{2}\mathbf{H}_{1}}{%
\partial (C^{1})^{2}}=0, \\
&& \\
\frac{\partial \mathbf{H}_{0}}{\partial C^{0}}\frac{\partial ^{2}\mathbf{H}%
_{1}}{\partial (C^{0})^{2}} &=&\frac{\partial \mathbf{H}_{1}}{\partial C^{0}}%
\frac{\partial ^{2}\mathbf{H}_{0}}{\partial (C^{0})^{2}}+2C^{1}\frac{%
\partial ^{2}\mathbf{H}_{0}}{\partial (C^{0})^{2}}\frac{\partial ^{2}\mathbf{%
H}_{1}}{\partial C^{0}\partial C^{1}},
\end{eqnarray*}%
whose solutions are the \textit{same} as solutions of the system (\textbf{%
\ref{fun}}).

\subsection{The third integrability criterion (the Hamiltonian hydrodynamic
reductions)}

Suppose the Hamiltonian hydrodynamic chain (\textbf{\ref{ak}}) is
integrable. Then its Hamiltonian hydrodynamic reduction (\textbf{\ref{gam}})
must be consistent with the generating function of conservation laws (%
\textbf{\ref{3}}), which is a particular case of more general generating
function of conservation laws $s_{t}=\partial _{x}(V(s)\upsilon )$
considered in details in \textbf{\cite{Maks+vech}}. The compatibility
condition $\partial _{i}(\partial _{k}s)=\partial _{k}(\partial _{i}s)$
(written via flat coordinates $c^{k}$; see (\textbf{\ref{reda}})) yields the
ODE%
\begin{equation*}
\gamma \mathbf{h}_{0}^{\prime }\mathbf{h}_{0}^{\prime \prime \prime
}=(3\gamma -1)\mathbf{h}_{0}^{\prime \prime ^{2}},
\end{equation*}%
whose solution $\mathbf{h}_{0}=(C^{0})^{\frac{1-\gamma }{1-2\gamma }}$ again
(see (\textbf{\ref{fun}})), where%
\begin{equation*}
\partial _{i}s=\frac{\varepsilon _{i}(c^{i})^{\frac{\gamma }{\gamma -1}}s^{%
\frac{\gamma }{\gamma -1}}}{(c^{i})^{\frac{1}{\gamma -1}}-s^{\frac{1}{\gamma
-1}}}\left( \frac{\gamma }{\gamma -1}\frac{\mathbf{h}_{0}^{\prime }}{\mathbf{%
h}_{0}^{\prime \prime }}+\sum \frac{\varepsilon _{k}(c^{k})^{\frac{2\gamma }{%
\gamma -1}}}{(c^{k})^{\frac{1}{\gamma -1}}-s^{\frac{1}{\gamma -1}}}\right)
^{-1}.
\end{equation*}%
Since $\partial _{i}s=-\partial _{i}\lambda /\partial _{s}\lambda $ (see 
\textbf{\cite{Maks+vec}}), the equation of the Riemann surface $\lambda (%
\mathbf{c};s)$ can be found in quadratures%
\begin{equation*}
d\lambda =s^{-3}\sum \frac{\varepsilon _{k}(c^{k})^{\frac{2\gamma -1}{\gamma
-1}}}{(c^{k})^{\frac{1}{\gamma -1}}-s^{\frac{1}{\gamma -1}}}ds-s^{-2}\sum 
\frac{\varepsilon _{k}(c^{k})^{\frac{\gamma }{\gamma -1}}}{(c^{k})^{\frac{1}{%
\gamma -1}}-s^{\frac{1}{\gamma -1}}}dc^{k}.
\end{equation*}%
Taking into account the moment decomposition (\textbf{\ref{reda}}), the
equation of the Riemann surface can be written in the form%
\begin{equation}
\lambda =\underset{k=0}{\overset{\infty }{\sum }}C^{k}s^{\frac{1-2\gamma -k}{%
\gamma -1}}.  \label{m}
\end{equation}%
Thus, this equation of the Riemann mapping is valid for whole hydrodynamic
chain (as well as for \textit{any} its hydrodynamic reduction). Indeed,

\textbf{Theorem 3}: \textit{The Gibbons equation}%
\begin{equation}
\lambda _{t^{1}}-\frac{\gamma (1-\gamma )}{1-2\gamma }s^{\frac{1}{\gamma -1}%
}(C^{0})^{\frac{\gamma }{1-2\gamma }}\lambda _{x}=\frac{\partial \lambda }{%
\partial s}\left[ s_{t^{1}}+\frac{(1-\gamma )^{2}}{1-2\gamma }\partial
_{x}(s^{\frac{\gamma }{\gamma -1}}(C^{0})^{\frac{\gamma }{1-2\gamma }})%
\right]   \label{azh}
\end{equation}%
\textit{connects the Hamiltonian hydrodynamic chain} (\textbf{\ref{j}}) 
\textit{with the Riemann mapping} (\textbf{\ref{m}}) ($\gamma \neq 1$, $%
\gamma \neq 1/2$).

\textbf{Proof}: can be obtained by the direct substitution (\textbf{\ref{j}}%
) and (\textbf{\ref{m}}) in (\textbf{\ref{azh}}).

\textbf{Remark}: The Gibbons equation

\begin{equation*}
\lambda _{t^{1}}-\frac{e^{s}}{C^{0}}\lambda _{x}=\frac{\partial \lambda }{%
\partial s}\left[ s_{t^{1}}-\partial _{x}(\frac{e^{s}}{C^{0}})\right] 
\end{equation*}%
connects the Hamiltonian hydrodynamic chain (\textbf{\ref{4}}) with the
Riemann mapping%
\begin{equation*}
\lambda =\underset{k=0}{\overset{\infty }{\sum }}C^{k}e^{-(k+1)s}.
\end{equation*}%
The Gibbons equation

\begin{equation*}
\lambda _{t^{1}}-\frac{e^{C^{0}}}{2s^{2}}\lambda _{x}=\frac{\partial \lambda 
}{\partial s}\left[ s_{t^{1}}+\partial _{x}\left( \frac{e^{C^{0}}}{2s}%
\right) \right] 
\end{equation*}%
connects the Hamiltonian hydrodynamic chain (\textbf{\ref{5}}) with the
Riemann mapping%
\begin{equation*}
\lambda =-\ln s+\underset{k=0}{\overset{\infty }{\sum }}C^{k}s^{2k}.
\end{equation*}

Let us consider two hydrodynamic chains%
\begin{equation*}
C_{t}^{k}=\underset{n=0}{\overset{k+1}{\sum }}F_{n}^{k}(\mathbf{C})C_{x}^{n}%
\text{, \ \ \ \ \ \ \ }\tilde{C}_{t}^{k}=\underset{n=0}{\overset{k+1}{\sum }}%
\tilde{F}_{n}^{k}(\mathbf{\tilde{C}})\tilde{C}_{x}^{n}\text{,}
\end{equation*}%
related by the infinitely many \textit{invertible} transformations%
\begin{equation}
\tilde{C}^{k}=\tilde{C}^{k}(C^{0},C^{1},...,C^{k})\text{, \ \ \ \ \ \ }%
k=0,1,2,...  \label{ku}
\end{equation}

\textbf{Theorem 4}: \textit{The hydrodynamic chain}%
\begin{equation}
\partial _{t}C_{(\alpha )}^{k}=\frac{\gamma }{\alpha }(C_{(\alpha
)}^{0})^{\gamma /\alpha -1}[\alpha C_{(\alpha )}^{0}\partial _{x}C_{(\alpha
)}^{k+1}+(k+1-\alpha )C_{(\alpha )}^{k+1}\partial _{x}C_{(\alpha )}^{0}]
\label{fi}
\end{equation}%
\textit{is equivalent to the Hamiltonian hydrodynamic chain} (\textbf{\ref{j}%
})%
\begin{equation*}
C_{t}^{k}=\frac{\gamma }{1-2\gamma }(C^{0})^{\frac{3\gamma -1}{1-2\gamma }%
}[(1-2\gamma )C^{0}C_{x}^{k+1}+(k+2\gamma )C^{k+1}C_{x}^{0}],
\end{equation*}%
\textit{where} $C^{k}\equiv C_{(1-2\gamma )}^{k}$.

\textbf{Proof}: The hydrodynamic chain (\textbf{\ref{fi}}) satisfies the
Gibbons equation (\textbf{\ref{azh}})%
\begin{equation*}
\lambda _{t}-\gamma s^{\frac{1}{\gamma -1}}(C^{0})^{\gamma /\alpha }\lambda
_{x}=\frac{\partial \lambda }{\partial s}[s_{t}+(\gamma -1)\partial _{x}(s^{%
\frac{\gamma }{\gamma -1}}(C^{0})^{\gamma /\alpha })],
\end{equation*}%
where $s=p^{\gamma -1}$ and the Riemann mapping (cf. (\textbf{\ref{m}})) is
given by%
\begin{equation*}
\lambda =\underset{k=0}{\overset{\infty }{\sum }}C_{(\alpha )}^{k}p^{\alpha
-k}\equiv \left[ \underset{k=0}{\overset{\infty }{\sum }}C^{k}p^{1-2\gamma
-k}\right] ^{\frac{\alpha }{1-2\gamma }}.
\end{equation*}%
Thus, the invertible transformations (\textbf{\ref{ku}}) can be derived from
the above formula.

\textbf{Remark}: The hydrodynamic chain (\textbf{\ref{ak}}) is a particular
case of three parametric hydrodynamic chain (see \textbf{\cite{Maks+vec}})%
\begin{equation*}
C_{t}^{k}=\mathbf{H}_{0}^{\prime }(C^{0})C_{x}^{k+1}+[(\alpha
k+2)C^{k+1}+\beta kC^{k}+\gamma kC^{k-1}]\partial _{x}\mathbf{H}_{0}^{\prime
}(C^{0}),
\end{equation*}%
determined by the Poisson bracket%
\begin{equation*}
\{C^{k},C^{n}\}=[B_{k,n}\partial _{x}+\partial _{x}B_{n,k}]\delta
(x-x^{\prime }),
\end{equation*}%
where%
\begin{equation*}
B_{k,n}=(\alpha k+1)A^{k+n+1}+\beta kA^{k+n}+\gamma kA^{k+n-1}.
\end{equation*}%
This Poisson bracket looks more complicated than (\textbf{\ref{dorf}}). The
moment decomposition%
\begin{equation*}
C^{k}=\sum \int V_{i}^{\prime ^{k}}dz_{i},
\end{equation*}%
where the $V(s)$ satisfies the ODE%
\begin{equation*}
2VV^{\prime \prime }=\alpha V^{\prime ^{2}}+\beta V^{\prime }+\gamma
\end{equation*}%
is more complicated than (\textbf{\ref{reda}}). However, a complexity of the
generating function of conservation laws (see \textbf{\cite{Maks+vec}}) is
the same as in the above example (\textbf{\ref{3}})%
\begin{equation*}
s_{t}=\partial _{x}[\mathbf{h}^{\prime }(\Delta )z^{\prime }(s)],
\end{equation*}%
where the Hamiltonian density can be of three types $\mathbf{h}(\Delta
)=\exp \Delta $, $\mathbf{h}(\Delta )=\ln \Delta $, $\mathbf{h}(\Delta
)=\Delta ^{\varepsilon }$ and function $z(s)$ is given in the implicit form%
\begin{equation*}
s=\int \frac{dz}{\sqrt{2V(z)}}.
\end{equation*}

\section{Hydrodynamic chains associated with the Kupershmidt brackets}

The method of hydrodynamic reductions was established in \textbf{\cite%
{Gib+Tsar}} (for a description of solutions of the Benney hydrodynamic
chain) and developed in \textbf{\cite{Fer+Kar}} (see also \textbf{\cite%
{FerKarMax}}) for a classification of 2+1 quasilinear systems. In this paper
this method is applied to classification of integrable hydrodynamic chains
(see \textbf{\cite{Kuper}}). The second example is given by the Hamiltonian
hydrodynamic chain%
\begin{equation*}
B_{t^{1}}^{k}=(1+\beta )\frac{\partial \mathbf{H}_{1}}{\partial B^{1}}%
B_{x}^{k+1}+\beta \frac{\partial \mathbf{H}_{1}}{\partial B^{0}}%
B_{x}^{k}+(k+1+2\beta )B^{k+1}\left( \frac{\partial \mathbf{H}_{1}}{\partial
B^{1}}\right) _{x}+(k+2\beta )B^{k}\left( \frac{\partial \mathbf{H}_{1}}{%
\partial B^{0}}\right) _{x}\text{, \ \ \ }k=0,1,...
\end{equation*}%
associated with the Kupershmidt Poisson bracket%
\begin{equation}
\{B^{k},B^{n}\}=[(k+\beta )B^{k+n}\partial _{x}+(n+\beta )\partial
_{x}B^{k+n}]\delta (x-x^{\prime }).  \label{kuper}
\end{equation}%
If the Hamiltonian density $\mathbf{H}_{1}$ is an arbitrary, then this
hydrodynamic chain has two conservation laws only. The conservation law of
the momentum%
\begin{equation}
B_{t^{1}}^{0}=\partial _{x}[(1+2\beta )B^{1}\frac{\partial \mathbf{H}_{1}}{%
\partial B^{1}}+2\beta B^{0}\frac{\partial \mathbf{H}_{1}}{\partial B^{0}}%
-\beta \mathbf{H}_{1}]  \label{mom}
\end{equation}%
and the conservation law of the energy%
\begin{equation*}
\partial _{t^{1}}\mathbf{H}_{1}=\partial _{x}\left[ (1+\beta )B^{2}\left( 
\frac{\partial \mathbf{H}_{1}}{\partial B^{1}}\right) ^{2}+(1+2\beta )B^{1}%
\frac{\partial \mathbf{H}_{1}}{\partial B^{0}}\frac{\partial \mathbf{H}_{1}}{%
\partial B^{1}}+\beta B^{0}\left( \frac{\partial \mathbf{H}_{1}}{\partial
B^{0}}\right) ^{2}\right] .
\end{equation*}

Taking into account that this Poisson bracket is reducible by the moment
decomposition (\textbf{\ref{simple}}) (where $\beta \rightarrow 1/\beta $, $%
\gamma =2$)%
\begin{equation}
B^{k}=\frac{\beta }{k+2\beta }\underset{i=1}{\overset{N}{\sum }}\varepsilon
_{i}(b^{i})^{k/\beta +2},  \label{kum}
\end{equation}%
the Hamiltonian hydrodynamic chain (see (\textbf{\ref{kuper}}))%
\begin{equation}
B_{t^{m}}^{k}=[(k+\beta )B^{k+n}\partial _{x}+(n+\beta )\partial _{x}B^{k+n}]%
\frac{\delta \mathbf{\bar{H}}_{m}}{\delta B^{n}}\text{, \ \ \ \ \ }%
k=0,1,2,...  \label{kun}
\end{equation}%
reduces to the hydrodynamic type system%
\begin{equation}
b_{t^{m}}^{i}=\beta \partial _{x}\left( (b^{i})^{1+n/\beta }\frac{\partial 
\mathbf{h}_{m}}{\partial B^{n}}\right)  \label{kuppa}
\end{equation}%
where $\mathbf{\bar{H}}_{m}\mathbf{=}\int \mathbf{H}%
_{m}(B^{0},B^{1},...,B^{m})dx$ and $\mathbf{h}_{m}(\mathbf{b})=\mathbf{H}%
_{m}|_{B^{k}=B^{k}(\mathbf{b})}$. Thus, a description of all admissible
Hamiltonian densities $\mathbf{H}_{m}$ can be derived from the consistency
of the above hydrodynamic type system and the generating function of
conservation laws%
\begin{equation}
q_{t^{m}}=\beta \partial _{x}\left( q^{1+n/\beta }\frac{\partial \mathbf{h}%
_{m}}{\partial B^{n}}\right) .  \label{kupo}
\end{equation}%
Then the above hydrodynamic type system (\textbf{\ref{kuppa}}) in the
Riemann invariants has the form%
\begin{equation*}
r_{t^{m}}^{i}=(n+\beta )(q^{i})^{n/\beta }\frac{\partial \mathbf{h}_{m}}{%
\partial B^{n}}r_{x}^{i}.
\end{equation*}%
In general case the Gibbons--Tsarev system can be derived from the
compatibility conditions $\partial _{t^{i}}(\partial _{t^{j}}q)=\partial
_{t^{j}}(\partial _{t^{i}}q)$, where $\partial _{i}\equiv \partial /\partial
r^{i}$ (cf. (\textbf{\ref{tri}}))%
\begin{equation*}
\partial _{i}q=\frac{\beta q^{1+n/\beta }\partial _{i}(\frac{\partial 
\mathbf{h}_{m}}{\partial B^{n}})}{(k+\beta )[(q^{i})^{k/\beta }-q^{k/\beta }]%
\frac{\partial \mathbf{h}_{m}}{\partial B^{k}}},
\end{equation*}%
and from the compatibility conditions $\partial _{i}(\partial
_{k}B^{n})=\partial _{k}(\partial _{i}B^{n})$, where $i\neq k$ and $%
n=1,2,...,m$.

Without lost of generality let us restrict our consideration on the
Hamiltonian density $\mathbf{H}_{1}(B^{0},B^{1})$. Then%
\begin{equation*}
\partial _{i}q=\frac{\beta }{\beta +1}\frac{q}{\partial \mathbf{h}%
_{1}/\partial B^{1}}\frac{\partial _{i}(\partial \mathbf{h}_{1}/\partial
B^{0})+q^{1/\beta }\partial _{i}(\partial \mathbf{h}_{1}/\partial B^{1})}{%
(q^{i})^{1/\beta }-q^{1/\beta }}.
\end{equation*}

The last step \textit{before} a computation of the compatibility conditions $%
\partial _{i}(\partial _{k}q)=\partial _{k}(\partial _{i}q)$ is the
substitution the link (into the above expression for each $\partial _{i}q$)%
\begin{equation*}
\partial _{i}B^{1}=\frac{(\beta +1)(q^{i})^{1/\beta }\frac{\partial \mathbf{h%
}_{1}}{\partial B^{1}}-(2\beta +1)B^{1}\frac{\partial ^{2}\mathbf{h}_{1}}{%
\partial B^{0}\partial B^{1}}-2\beta B^{0}\frac{\partial ^{2}\mathbf{h}_{1}}{%
\partial (B^{0})^{2}}}{(\beta +1)\frac{\partial \mathbf{h}_{1}}{\partial
B^{1}}+(2\beta +1)B^{1}\frac{\partial ^{2}\mathbf{h}_{1}}{\partial
(B^{1})^{2}}+2\beta B^{0}\frac{\partial ^{2}\mathbf{h}_{1}}{\partial
B^{0}\partial B^{1}}}\partial _{i}B^{0},
\end{equation*}%
which is a consequence on the first equation of the hydrodynamic chain (see (%
\textbf{\ref{mom}})).

The compatibility conditions $\partial _{i}(\partial _{k}B^{1})=\partial
_{k}(\partial _{i}B^{1})$ imply a last restriction on the dependence of the
Hamiltonian density $\mathbf{h}_{1}$ on $B^{0}$ and $B^{1}$.

The same Gibbons--Tsarev system but written via field variables $b^{k}$ (see
(\textbf{\ref{kuppa}})) can be derived from the compatibility conditions $%
\partial _{i}(\partial _{j}q)=\partial _{j}(\partial _{i}q)$, where $%
\partial _{i}\equiv \partial /\partial b^{i}$ and%
\begin{equation*}
\partial _{i}q=\frac{\beta }{\beta +1}\frac{(\sum (b^{k})^{1+1/\beta
}\partial _{k}q-q^{1+1/\beta })\partial _{i}(\partial \mathbf{h}%
_{1}/\partial B^{1})+(\sum b^{k}\partial _{k}q-q)\partial _{i}(\partial 
\mathbf{h}_{1}/\partial B^{0})}{[q^{1/\beta }-(b^{i})^{1/\beta }]\partial 
\mathbf{h}_{1}/\partial B^{0}}.
\end{equation*}%
\textit{Before} a substitution of the above expression in the compatibility
conditions $\partial _{i}(\partial _{k}q)=\partial _{k}(\partial _{i}q)$,
one must compute $\Sigma (b^{k})^{1+1/\beta }\partial _{k}q$ and $\Sigma
b^{k}\partial _{k}q$. This investigation in details will be presented
elsewhere.

This is a particular case of more general problem: a description of all
admissible functions $\psi (B^{0},B^{1};q)$ (i.e. the classification of
integrable hydrodynamic chains) determined by the generating function of
conservation laws $q_{t}=\partial _{x}\psi (B^{0},B^{1};q)$ (see details in 
\textbf{\cite{Maks+vech}}). However, if the compatibility conditions $%
\partial _{i}(\partial _{k}q)=\partial _{k}(\partial _{i}q)$ imply the
extended Gibbons--Tsarev system describing all admissible functions $\psi
(B^{0},B^{1};q)$ and corresponding $N$ component reductions parameterized by 
$N$ arbitrary functions of a single variable, if the compatibility
conditions $\partial _{i}(\partial _{k}q)=\partial _{k}(\partial _{i}q)$ for
integrable 2+1 quasilinear systems like dKP (see \textbf{\cite{Gib+Tsar}}
and \textbf{\cite{Fer+Kar}}) imply the Gibbons--Tsarev system describing $N$
component reductions parameterized by $N$ arbitrary functions of a single
variable only, the compatibility conditions $\partial _{i}(\partial
_{k}q)=\partial _{k}(\partial _{i}q)$ for integrable hydrodynamic chains
connected with the Kupershmidt Poisson bracket \textit{reduce the
Gibbons--Tsarev system to the sole ODE} only. It means that the advantage of
this Hamiltonian hydrodynamic reductions approach in the application to the
Hamiltonian hydrodynamic chains, whose Poisson brackets are reducible, is
that the Gibbons--Tsarev system implies just a restriction on the
Hamiltonian density.

Integrable hydrodynamic chains associated with the Kupershmidt Poisson
bracket (\textbf{\ref{kuper}}) can be determined by the \textit{linear}
Hamiltonian density $\mathbf{H}_{1}=B^{1}+f(B^{0})$, the \textit{quasilinear}
Hamiltonian density $\mathbf{H}_{1}=g(B^{0})B^{1}+f(B^{0})$ and the \textit{%
fully nonlinear} Hamiltonian density $\mathbf{H}_{1}(B^{0},B^{1})$.

\subsection{The \textit{linear} Hamiltonian density}

The integrability of the hydrodynamic chain determined by the \textit{linear}
Hamiltonian density $\mathbf{H}_{1}=B^{1}+f(B^{0})$%
\begin{equation*}
B_{t^{1}}^{k}=B_{x}^{k+1}+\frac{\beta }{\beta +1}f^{\prime }(B^{0})B_{x}^{k}+%
\frac{k+2\beta }{1+\beta }B^{k}f^{\prime \prime }(B^{0})B_{x}^{0}\text{, \ \
\ }k=0,1,...
\end{equation*}%
can be examined with the aid of the method of hydrodynamic reductions, where
the generating function of conservation laws is given by (see (\textbf{\ref%
{kupo}}))%
\begin{equation}
q_{t^{1}}=\beta \partial _{x}\left( q^{1+1/\beta }+f^{\prime
}(B^{0})q\right) .  \label{qu}
\end{equation}%
In this case the Gibbons--Tsarev system can be derived from the
compatibility conditions $\partial _{i}(\partial _{k}q)=\partial
_{k}(\partial _{i}q)$ and $\partial _{i}(\partial _{k}B^{1})=\partial
_{k}(\partial _{i}B^{1})$, where $\partial _{k}\equiv \partial /\partial
r^{k}$ and%
\begin{equation*}
\partial _{i}q=\frac{\beta }{\beta +1}q\frac{f^{\prime \prime
}(B^{0})\partial _{i}B^{0}}{(q^{i})^{1/\beta }-q^{1/\beta }}\text{, \ \ \ \
\ }\partial _{i}B^{1}=[(q^{i})^{1/\beta }-\frac{2\beta }{\beta +1}%
B^{0}f^{\prime \prime }(B^{0})]\partial _{i}B^{0}.
\end{equation*}%
The function $f(B^{0})$ also can be found from the compatibility conditions $%
\partial _{i}(\partial _{k}q)=\partial _{k}(\partial _{i}q)$, where $%
\partial _{k}\equiv \partial /\partial b^{k}$ and%
\begin{equation*}
\partial _{i}q=\frac{q}{q^{1/\beta }-(b^{i})^{1/\beta }}\left[ \sum \frac{%
b^{k}}{q^{1/\beta }-(b^{k})^{1/\beta }}-\frac{\beta +1}{\beta }\frac{%
f^{\prime }(B^{0})}{f^{\prime \prime }(B^{0})}\right] ^{-1}.
\end{equation*}%
The computational result is $f^{\prime \prime \prime }(B^{0})=0$. Since $%
B^{0}$ is a \textit{momentum} density (see (\textbf{\ref{mom}})), one can
choose $f(B^{0})=\gamma (B^{0})^{2}$. Thus, the \textit{linear} Hamiltonian
density $\mathbf{H}_{1}=B^{1}+\gamma (B^{0})^{2}$ determines the integrable
hydrodynamic chain%
\begin{equation}
B_{t^{1}}^{k}=B_{x}^{k+1}+\frac{2\beta \gamma }{\beta +1}B^{0}B_{x}^{k}+2%
\gamma \frac{k+2\beta }{1+\beta }B^{k}B_{x}^{0}\text{, \ \ \ }k=0,1,...,
\label{ru}
\end{equation}%
whose generating function of conservation laws is given by (see (\textbf{\ref%
{qu}}))%
\begin{equation}
q_{t^{1}}=\beta \partial _{x}\left( q^{1+1/\beta }+2\gamma B^{0}q\right) .
\label{qup}
\end{equation}%
This is exactly the Kupershmidt hydrodynamic chain (see \textbf{\cite{Kuper}}%
) up to scaling $B^{k}\rightarrow \frac{\beta +1}{2\gamma }B^{k}$.

\subsection{The Miura type and reciprocal transformations}

Let us consider two hydrodynamic chains%
\begin{equation*}
C_{t}^{k}=\underset{n=0}{\overset{k+1}{\sum }}F_{n}^{k}(\mathbf{C})C_{x}^{n}%
\text{, \ \ \ \ \ \ }B_{t}^{k}=\underset{n=0}{\overset{k+1}{\sum }}G_{n}^{k}(%
\mathbf{B})B_{x}^{n}\text{, \ \ \ \ \ }k=0,1,2,...,
\end{equation*}%
related by the infinitely many transformations%
\begin{equation}
B^{k}=B^{k}(C^{0},C^{1},...,C^{k+1})\text{, \ \ \ \ \ }k=0,1,2,...
\label{ura}
\end{equation}

\textbf{Definition 3}: \textit{The transformations} (\textbf{\ref{ura}}) 
\textit{connecting} $B-$\textit{hydrodynamic chain (right) and \textbf{%
modified}} $C-$\textit{hydrodynamic chain (left) are said to be the \textbf{%
Miura type transformations}}.

\textbf{Theorem 5}: \textit{The hydrodynamic chain} (\textbf{\ref{j}}) 
\textit{is connected with the Kupershmidt hydrodynamic chain}%
\begin{equation}
B_{y}^{k}=B_{z}^{k+1}+(\gamma -1)B^{0}B_{z}^{k}+(k+2\gamma )B^{k}B_{z}^{0}%
\text{, \ \ \ \ \ }k=0,1,2,...  \label{zu}
\end{equation}%
\textit{by the reciprocal transformation}%
\begin{equation*}
dz=\mathbf{H}_{0}(C^{0})dx+\gamma \mathbf{H}_{0}^{\prime
^{2}}(C^{0})C^{1}dt^{1}\text{, \ \ \ \ \ \ }dy^{1}=\frac{\gamma (1-\gamma )}{%
1-2\gamma }dt^{1},
\end{equation*}%
\textit{where} $(1-2\gamma )B^{k}=C^{k+1}(C^{0})^{-1+(k+1)/(1-2\gamma )}$
(see (\textbf{\ref{ura}})).

\textbf{Proof}: Can be obtained by a straightforward calculation.

\textbf{Remark}: The generating function of conservation laws (\textbf{\ref%
{3}}) transforms in the generating function of conservation laws for the
Kupershmidt hydrodynamic chains (\textbf{\ref{zu}})%
\begin{equation*}
q_{y}=\frac{\gamma -1}{\gamma }\partial _{z}[(1-2\gamma )q^{\frac{\gamma }{%
\gamma -1}}+\gamma B^{0}q],
\end{equation*}%
where the \textit{generating function of the Miura type transformations} is $%
s=\mathbf{H}_{0}(C^{0})q$.

Thus, if hydrodynamic reductions of the Kupershmidt hydrodynamic chains ($M-$%
dMKP hierarchy, see \textbf{\cite{Blaszak}}\ and \textbf{\cite{Manasa}}) are
known, then they can be recalculated to hydrodynamic reductions of $M-$dDym
hierarchy (see also details in \textbf{\cite{Szab}}) by the above reciprocal
transformation.

The hydrodynamic chains (\textbf{\ref{ru}}), (\textbf{\ref{zu}}) are
particular cases of the Kupershmidt hydrodynamic chain (see \textbf{\cite%
{Kuper}})%
\begin{equation}
B_{t}^{k}=B_{x}^{k+1}+\frac{1}{\varepsilon }B^{0}B_{x}^{k}+(k+\delta
)B^{k}B_{x}^{0}\text{, \ \ \ \ \ \ }k=0,1,2,...,  \label{kupu}
\end{equation}%
which are equivalent to each other under an invertible transformation for
any fixed value of the parameter $\varepsilon $ (see the theorem below).

\textbf{Theorem 6}: \textit{The Gibbons equation}%
\begin{equation}
\lambda _{t}-(p^{\varepsilon }+\frac{B^{0}}{\varepsilon })\lambda _{x}=\frac{%
\partial \lambda }{\partial p}\left[ p_{t}-\partial _{x}\left( \frac{%
p^{\varepsilon +1}}{\varepsilon +1}+\frac{B^{0}}{\varepsilon }p\right) %
\right]   \label{rima}
\end{equation}%
\textit{\ describes a deformation of the Riemann mapping}%
\begin{equation*}
\lambda =p^{\varepsilon (1-\delta )}+(1-\delta )\underset{k=0}{\overset{%
\infty }{\sum }}B^{k}p^{-\varepsilon (k+\delta )},
\end{equation*}%
\textit{where the coefficients }$B^{k}$ \textit{satisfy the Kupershmidt
hydrodynamic chain} (\textbf{\ref{kupu}})\textit{.}

\subsection{Reciprocal transformations and generating functions of
conservation laws}

The classification of integrable hydrodynamic chains associated with the
Kupershmidt Poisson brackets and determined by the \textit{quasilinear} and 
\textit{fully nonlinear} Hamiltonian densities will be given in separate
publication. In this sub-section we restrict our consideration on
corresponding generating functions of conservation laws.

\textbf{Theorem 7}: \textit{In the \textbf{quasilinear} case the generating
function of conservation laws (see} (\textbf{\ref{kupo}}))%
\begin{equation}
q_{t^{1}}=\beta \partial _{x}\left( g(B^{0})q^{1+1/\beta }+[g^{\prime
}(B^{0})B^{1}+f^{\prime }(B^{0})]q\right)   \label{v}
\end{equation}%
\textit{can be reduced to the \textbf{linear} case iff}%
\begin{equation*}
g(B^{0})=(B^{0}+\delta )^{-(1+1/\beta )}.
\end{equation*}

\textbf{Proof}: An arbitrary reciprocal transformation%
\begin{equation}
dz=Fdx+Gdt\text{, \ \ \ \ \ \ }dy=dt  \label{re}
\end{equation}%
reduces (\textbf{\ref{v}}) to%
\begin{equation}
p_{y^{1}}=\beta \partial _{z}\left( g(B^{0})F^{1+1/\beta }p^{1+1/\beta
}+[(g^{\prime }(B^{0})B^{1}+f^{\prime }(B^{0}))F-G/\beta ]p\right) ,
\label{w}
\end{equation}%
where the generating function of the Miura type transformations is $q=Fp$.
Since in the \textit{linear} case (see (\textbf{\ref{qup}}))%
\begin{equation}
p_{y^{1}}=\beta \partial _{z}\left( q^{1+1/\beta }+\tilde{B}^{0}p\right) ,
\label{wu}
\end{equation}%
then $g(B^{0})F^{1+1/\beta }=1$, i.e. the function $F$ must be a some
function of $B^{0}$. However, $F$ is a conservation law density, $B^{0}$ is
a conservation law density (see (\textbf{\ref{mom}})), then $F$ is a \textit{%
linear} function with respect to $B^{0}$. Thus, indeed, $g(B^{0})=(B^{0}+%
\delta )^{-(1+1/\beta )}$, where $\delta $ is an arbitrary constant.

The first Miura type transformation is (cf. (\textbf{\ref{w}}) and (\textbf{%
\ref{wu}}))%
\begin{equation*}
\tilde{B}^{0}=-2\delta \frac{\beta +1}{\beta }(B^{0}+\delta )^{-(2+1/\beta
)}B^{1}+(\delta -B^{0})f^{\prime }(B^{0})+f(B^{0}).
\end{equation*}%
Since $\tilde{B}^{0}$ is a conservation law density (see (\textbf{\ref{qup}}%
)) in the independent variables $z,y$, the r.h.s. of the expression%
\begin{equation*}
\tilde{B}^{0}(B^{0}+\delta )=-2\delta \frac{\beta +1}{\beta }(B^{0}+\delta
)^{-(1+1/\beta )}B^{1}+[(\delta -B^{0})f^{\prime
}(B^{0})+f(B^{0})](B^{0}+\delta )
\end{equation*}%
is a conservation law density in the independent variables $x,t$. However,
this r.h.s. must coincide with the \textit{quasilinear} Hamiltonian density $%
\mathbf{H}_{1}=g(B^{0})B^{1}+f(B^{0})$ up to the momentum density $B^{0}$
and an arbitrary constant $\varepsilon $. Thus, we have the ODE of the first
order%
\begin{equation*}
-2\delta \frac{\beta +1}{\beta }f(B^{0})=[(\delta -B^{0})f^{\prime
}(B^{0})+f(B^{0})](B^{0}+\delta ),
\end{equation*}%
whose solution is%
\begin{equation*}
f(B^{0})=(B^{0}-\delta )^{2+1/\beta }(B^{0}+\delta )^{-(1+1/\beta )}.
\end{equation*}%
Then the \textit{quasilinear} Hamiltonian density ($\xi $ is an arbitrary
constant)%
\begin{equation*}
\mathbf{H}_{1}=(B^{0}+\delta )^{-(1+1/\beta )}[B^{1}+\xi (B^{0}-\delta
)^{2+1/\beta }]
\end{equation*}%
creates an integrable hydrodynamic chain reducible to the Kupershmidt
hydrodynamic chain.

\textbf{Theorem 8}: \textit{In the \textbf{fully nonlinear} case the
generating function of conservation laws}%
\begin{equation}
q_{t^{1}}=\beta \partial _{x}\left( q^{1+1/\beta }\frac{\partial \mathbf{H}%
_{1}}{\partial B^{1}}+q\frac{\partial \mathbf{H}_{1}}{\partial B^{0}}\right)
\label{fully}
\end{equation}%
\textit{can be reduced to the \textbf{linear} case iff}%
\begin{equation}
\mathbf{H}_{1}=[B^{1}+\varphi (B^{0})]^{\frac{\beta }{2\beta +1}},
\label{sol}
\end{equation}%
\textit{where} $\varphi (B^{0})$ \textit{is a some function}.

\textbf{Proof}: An arbitrary reciprocal transformation (\textbf{\ref{re}})
reduces (\textbf{\ref{fully}}) to%
\begin{equation*}
p_{y^{1}}=\beta \partial _{z}\left[ F^{1+1/\beta }\frac{\partial \mathbf{H}%
_{1}}{\partial B^{1}}p^{1+1/\beta }+\left( F\frac{\partial \mathbf{H}_{1}}{%
\partial B^{0}}-\frac{G}{\beta }\right) p\right] .
\end{equation*}%
This generating function coincides with the \textit{linear} case if $%
F^{1+1/\beta }\partial \mathbf{H}_{1}/\partial B^{1}=1$. Since $\mathbf{H}%
_{1}$ depends on $B^{0}$ and $B^{1}$, then the conservation law density $%
F(B^{0},B^{1})$ must be a \textit{linear} function with respect to the
Hamiltonian density $\mathbf{H}_{1}$ up to the momentum density $B^{0}$ and
an arbitrary constant $\delta $, which can be removed by shift $z\rightarrow
z+\varepsilon y$, where $\varepsilon $ is an appropriate constant. Thus, we
have the ODE%
\begin{equation*}
(\mathbf{H}_{1})^{1+1/\beta }\frac{\partial \mathbf{H}_{1}}{\partial B^{1}}%
=1,
\end{equation*}%
whose solution is (\textbf{\ref{sol}}) up to an insufficient factor. The
function $\varphi (B^{0})$ can be determined in the same way as in the 
\textit{quasilinear} case, but we avoid this complicated computation in this
paper. Further details can be found in \textbf{\cite{FerMarMax}}.
Nevertheless, we would like to emphasize that an application of reciprocal
transformations to the quasilinear case and moreover to fully nonlinear case
significantly simplifies computations made by the Hamiltonian hydrodynamic
reductions method (see the beginning of this section).

\section{Hydrodynamic chains determined by the Kupershmidt--Manin bracket}

In this paper we restrict our consideration on the \textit{linear} case $%
\mathbf{H}_{2}=A^{2}/2+f(A^{0}$, $A^{1})$ in general. The hydrodynamic chain
(\textbf{\ref{b}})%
\begin{equation}
A_{t}^{k}=A_{x}^{k+1}+\frac{\partial f}{\partial A^{1}}A_{x}^{k}+(k+1)A^{k}%
\left( \frac{\partial f}{\partial A^{1}}\right) _{x}+kA^{k-1}\left( \frac{%
\partial f}{\partial A^{0}}\right) _{x}.  \label{lin}
\end{equation}%
creates the generating function of conservation laws (\textbf{\ref{genn}})%
\begin{equation}
p_{t}=\partial _{x}\left( \frac{p^{2}}{2}+\frac{\partial f}{\partial A^{1}}p+%
\frac{\partial f}{\partial A^{0}}\right) ,  \label{ser}
\end{equation}%
where $p$ is a some function of all moments $A^{0},A^{1},A^{2}$, ... and
``spectral parameter'' $\lambda $, which is not introduced yet. If $%
p\rightarrow \infty $, the corresponding flux of the generating function of
conservation laws (\textbf{\ref{ser}}) also tends to infinity. One can seek $%
p$ as the Laurent series in the form (\textbf{\ref{series}})%
\begin{equation}
p=\lambda -\mathbf{\tilde{H}}_{-1}-\frac{\mathbf{\tilde{H}}_{0}}{\lambda }-%
\frac{\mathbf{\tilde{H}}_{1}}{\lambda ^{2}}-\frac{\mathbf{\tilde{H}}_{2}}{%
\lambda ^{3}}-...  \label{fuk}
\end{equation}%
If hydrodynamic chain (\textbf{\ref{lin}}) is integrable, then the \textit{%
infinite} series of conservation laws%
\begin{eqnarray}
\partial _{t}\mathbf{\tilde{H}}_{-1} &=&\partial _{x}\left( \mathbf{\tilde{H}%
}_{0}+\frac{1}{2}\mathbf{\tilde{H}}_{-1}^{2}-\frac{\partial f}{\partial A^{0}%
}\right) \text{, \ \ \ \ \ }\partial _{t}\mathbf{\tilde{H}}_{0}=\partial _{x}%
\mathbf{\tilde{H}}_{1}\text{,}  \notag \\
&&  \label{egor} \\
\partial _{t}\mathbf{\tilde{H}}_{k} &=&\partial _{x}\left( \mathbf{\tilde{H}}%
_{k+1}-\frac{1}{2}\underset{m=0}{\overset{k-1}{\sum }}\mathbf{\tilde{H}}_{m}%
\mathbf{\tilde{H}}_{k-1-m}\right) ,  \notag
\end{eqnarray}%
where%
\begin{equation*}
\mathbf{\tilde{H}}_{-1}=\frac{\partial f}{\partial A^{1}},
\end{equation*}%
can be derived by the substitution this series (\textbf{\ref{fuk}}) in (%
\textbf{\ref{ser}}).

Since $\partial f/\partial A^{1}$ is a conservation law density, then $%
\partial f/\partial A^{1}$ must be a \textit{linear} function of
conservation law densities $A^{0}$ and $A^{1}$ (the Casimir density and the
momentum density of the Kupershmidt--Manin bracket) only. Thus,%
\begin{equation*}
f=\alpha A^{0}A^{1}+\beta (A^{1})^{2}+\varphi (A^{0}),
\end{equation*}%
where $\alpha $ and $\beta $ are arbitrary constants (a linear term
proportional $A^{1}$ is removed by a linear change of the independent
variable $x\rightarrow x+t\limfunc{const}$), $\varphi (A^{0})$ is not
determined function yet (if $\beta \neq 0$, the constant $\alpha $ is
unessential and can be removed by the shift $\alpha A^{0}+2\beta
A^{1}\rightarrow 2\beta A^{1}$, see \textbf{\cite{Fer+Dav}} and \textbf{\cite%
{Maks+Egor}}).

Let us consider sub-cases depending on a different number of such constants.

\subsection{The Benney hydrodynamic chain}

In the simplest case $\mathbf{\tilde{H}}_{-1}=\limfunc{const}$, then the
generating function of conservation laws (\textbf{\ref{ser}}) is (the
central term $\mu \limfunc{const}$ can be removed by a linear change of the
independent variable $x\rightarrow x+t\limfunc{const}$)%
\begin{equation*}
\mu _{t}=\partial _{x}(\frac{\mu ^{2}}{2}+f^{\prime }(A^{0})).
\end{equation*}%
If the hydrodynamic chain (\textbf{\ref{lin}})%
\begin{equation*}
A_{t}^{k}=A_{x}^{k+1}+kA^{k-1}f^{\prime \prime }(A^{0})A_{x}^{0}
\end{equation*}%
is integrable, then a substitution of the Laurent series (\textbf{\ref{fuk}})%
\begin{equation*}
\mu =\lambda -\frac{\mathbf{H}_{0}}{\lambda }-\frac{\mathbf{H}_{1}}{\lambda
^{2}}-\frac{\mathbf{H}_{2}}{\lambda ^{3}}-...
\end{equation*}%
yields the infinite series of conservation laws (\textbf{\ref{egor}})%
\begin{equation*}
\partial _{t}\mathbf{H}_{0}=\partial _{x}\mathbf{H}_{1}\text{, \ \ \ \ \ \ \
\ \ \ }\partial _{t}\mathbf{H}_{k}=\partial _{x}\left( \mathbf{H}_{k+1}-%
\frac{1}{2}\underset{m=0}{\overset{k-1}{\sum }}\mathbf{H}_{m}\mathbf{H}%
_{k-1-m}\right) ,
\end{equation*}%
where%
\begin{equation*}
\mathbf{H}_{0}=f^{\prime }(A^{0}).
\end{equation*}%
Since $f^{\prime }(A^{0})$ is a conservation law density, then $f^{\prime
}(A^{0})$ must be a \textit{linear} function of the conservation law density 
$A^{0}$. Thus, one can choose $f(A^{0})=(A^{0})^{2}/2$ up to unessential
constants, which can be removed by the Galilean transformation, a shift and
a scaling. Thus, the first classification result is precisely the Benney
hydrodynamic chain (\textbf{\ref{bm}}).

\subsection{The \textit{mixed} modified Benney hydrodynamic chain}

Let us consider the most general case when all constants are not vanished.
In this case we replace the notation $A^{k}\rightarrow E^{k}$. Taking into
account that%
\begin{equation*}
E_{t}^{0}=\partial _{x}\left( E^{1}+E^{0}\frac{\partial f}{\partial E^{1}}%
\right) \text{,\ \ \ \ \ \ }E_{t}^{1}=\partial _{x}[E^{2}+2E^{1}\frac{%
\partial f}{\partial E^{1}}+E^{0}\frac{\partial f}{\partial E^{0}}-f],
\end{equation*}%
one can obtain%
\begin{equation*}
\mathbf{\tilde{H}}_{0}=2\beta E^{2}+4\beta ^{2}(E^{1})^{2}+4\alpha \beta
E^{0}E^{1}+\frac{\alpha ^{2}}{2}(E^{0})^{2}+2\alpha E^{1}+(1+2\beta
E^{0})\varphi ^{\prime }-2\beta \varphi 
\end{equation*}%
from the first conservation law (\textbf{\ref{egor}}). Since the Hamiltonian
density is $\mathbf{H}_{2}=E^{2}/2+\alpha E^{0}E^{1}+\beta
(E^{1})^{2}+\varphi (E^{0})$, then the conservation law density $\mathbf{%
\tilde{H}}_{0}-4\beta \mathbf{H}_{2}$ must be a \textit{linear} function of $%
E^{0}$ and $E^{1}$ only. It means, that $\mathbf{\tilde{H}}_{0}$ exists iff
the function $\varphi $ satisfies the ODE of the first order%
\begin{equation}
(1+2\beta E^{0})\varphi ^{\prime }-6\beta \varphi +\frac{\alpha ^{2}}{2}%
(E^{0})^{2}+\gamma E^{0}+\delta =0,  \label{prom}
\end{equation}%
where $\gamma $ and $\delta $ are arbitrary constants. Integrating this
equation, one can obtain the Hamiltonian%
\begin{equation*}
\mathbf{\bar{H}}_{2}=\int \left[ \frac{E^{2}}{2}+\alpha E^{0}E^{1}+\beta
(E^{1})^{2}-2\frac{\beta ^{2}\varepsilon ^{2}}{3}(E^{0})^{3}+(\frac{\alpha
^{2}}{4\beta }-\beta \varepsilon ^{2})(E^{0})^{2}\right] dx,
\end{equation*}%
where $\varepsilon $ is an arbitrary constant.

Thus, the most general (\textit{mixed modified} Benney) hydrodynamic chain%
\begin{equation*}
E_{t}^{k}=E_{x}^{k+1}+(\alpha E^{0}+2\beta E^{1})E_{x}^{k}+(k+1)E^{k}(\alpha
E^{0}+2\beta E^{1})_{x}+kE^{k-1}(\alpha E^{1}-2\beta ^{2}\varepsilon
^{2}(E^{0})^{2}+(\frac{\alpha ^{2}}{2\beta }-2\beta \varepsilon
^{2})E^{0})_{x}.
\end{equation*}%
determined by the Kupershmidt--Manin bracket and the \textit{linear}
Hamiltonian has the generating function of conservation laws (instead of $p$
we use $s$ for this case)%
\begin{equation*}
s_{t}=\partial _{x}\left( \frac{s^{2}}{2}+(\alpha E^{0}+2\beta
E^{1})s+\alpha E^{1}-2\beta ^{2}\varepsilon ^{2}(E^{0})^{2}+(\frac{\alpha
^{2}}{2\beta }-2\beta \varepsilon ^{2})E^{0}\right) .
\end{equation*}

\textbf{\ Remark}: \textit{The substitution}%
\begin{equation*}
\mu =s+\alpha E^{0}+2\beta E^{1}
\end{equation*}%
\textit{connects the above generating function of conservation laws with the
generating function of conservation laws}%
\begin{equation}
\mu _{t}=\partial _{x}\left( \frac{\mu ^{2}}{2}+A^{0}\right)   \label{bmg}
\end{equation}%
\textit{for the Benney hydrodynamic chain }(\textbf{\ref{bm}})\textit{.}

If $\varepsilon =0$, then the above hydrodynamic chain reduces to the 
\textit{twice modified} Benney hydrodynamic chain%
\begin{equation*}
C_{t}^{k}=C_{x}^{k+1}+(\alpha C^{0}+2\beta C^{1})C_{x}^{k}+(k+1)C^{k}(\alpha
C^{0}+2\beta C^{1})_{x}+kC^{k-1}(\alpha C^{1}+\frac{\alpha ^{2}}{2\beta }%
C^{0})_{x}
\end{equation*}%
connected with the generating function of conservation laws (instead of $p$
we use $q$ for this case)%
\begin{equation*}
q_{t}=\partial _{x}\left( \frac{q^{2}}{2}+(\alpha C^{0}+2\beta
C^{1})q+\alpha C^{1}+\frac{\alpha ^{2}}{2\beta }C^{0}\right) .
\end{equation*}

\textbf{Remark}: \textit{The substitution}%
\begin{equation*}
\mu =q+\alpha C^{0}+2\beta C^{1}
\end{equation*}%
\textit{connects the above generating function of conservation laws with the
generating function of conservation laws} (\textbf{\ref{bmg}}) \textit{for
the Benney hydrodynamic chain }(\textbf{\ref{bm}})\textit{, where}%
\begin{equation*}
\mathbf{\tilde{H}}_{0}=2\beta C^{2}+2\alpha C^{1}+\frac{\alpha ^{2}}{2\beta }%
C^{0}+\alpha ^{2}(C^{0})^{2}+4\alpha \beta C^{0}C^{1}+4\beta ^{2}(C^{1})^{2}.
\end{equation*}

If $\beta =0$, then the Hamiltonian (see (\textbf{\ref{prom}}))%
\begin{equation*}
\mathbf{\bar{H}}_{2}=\int \left( \frac{B^{2}}{2}+\alpha B^{0}B^{1}-\frac{%
\alpha ^{2}}{6}(B^{0})^{3}-\frac{\gamma }{2}(B^{0})^{2}\right) dx
\end{equation*}%
determines the \textit{modified} Benney hydrodynamic chain (the constant $%
\alpha $ can be removed by scaling $\alpha B^{k}\rightarrow B^{k}$, $\alpha
\neq 0$)%
\begin{equation*}
B_{t}^{k}=B_{x}^{k+1}+\alpha B^{0}B_{x}^{k}+\alpha
(k+1)B^{k}B_{x}^{0}+kB^{k-1}(\alpha B^{1}-\frac{\alpha ^{2}}{2}%
(B^{0})^{2}-\gamma B^{0})_{x}
\end{equation*}%
connected with the generating function of conservation laws%
\begin{equation*}
p_{t}=\partial _{x}\left( \frac{p^{2}}{2}+\alpha B^{0}p+\alpha B^{1}-\frac{%
\alpha ^{2}}{2}(B^{0})^{2}-\gamma B^{0}\right) .
\end{equation*}

\textbf{Remark}: \textit{The substitution}%
\begin{equation*}
p=\mu -\alpha B^{0}
\end{equation*}%
\textit{connects the above generating function of conservation laws with the
generating function of conservation laws}%
\begin{equation*}
\mu _{t}=\partial _{x}\left( \frac{\mu ^{2}}{2}+A^{0}\right) 
\end{equation*}%
\textit{for the Benney hydrodynamic chain }(\textbf{\ref{bm}})\textit{, where%
} $A^{0}=2\alpha B^{1}-\gamma B^{0}$.

\subsection{Reciprocal transformations and non-\textit{linear} cases}

In comparison with the previous section the \textit{quasilinear} case $%
\mathbf{H}_{2}=g(A^{0},A^{1})A^{2}+f(A^{0},A^{1})$ has \textit{two}
sub-cases. It is easy to see by an application of the reciprocal
transformation (\textbf{\ref{re}}) to the generating function of
conservation laws (\textbf{\ref{genn}})%
\begin{equation*}
p_{t}=\partial _{x}(g(A^{0},A^{1})p^{2}+\text{lower order terms}).
\end{equation*}%
The modified generating function of conservation laws is%
\begin{equation*}
\tilde{p}_{y}=\partial _{z}(\tilde{p}^{2}+\text{lower order terms}),
\end{equation*}%
where%
\begin{equation*}
g(A^{0},A^{1})F^{2}=1.
\end{equation*}%
If $g$ depends on $A^{0}$ only, then the conservation law density $F$ is a 
\textit{linear} function of the Casimir density $A^{0}$. Thus, $%
g=(A^{0}+\delta )^{-2}$. If $g$ depends on the both moments $A^{0}$ and $%
A^{1}$, then the conservation law density $F$ is a \textit{linear} function
of the Casimir and the momentum densities $A^{0}$ and $A^{1}$. Thus, $%
g=(A^{1}+\delta )^{-2}$ (a linear term proportional $A^{0}$ is removed by a
linear change of the independent variable $x\rightarrow x+t\limfunc{const}$;
see \textbf{\cite{Fer+Dav}}).

The \textit{fully nonlinear} case is reducible to the \textit{linear} case
under the reciprocal transformation (\textbf{\ref{re}}) if%
\begin{equation*}
\frac{\partial \mathbf{H}_{2}}{\partial A^{2}}F^{2}=1.
\end{equation*}%
Since $F$ is a conservation law density depending on all three moments $%
A^{0},A^{1},A^{2}$, then $F$ must be a \textit{linear} function of the
Hamiltonian density $\mathbf{H}_{2}$ up to the momentum and the Casimir
densities, which can be removed by a shift $x\rightarrow x+\varepsilon t$,
where $\varepsilon $ is an appropriate constant. Thus, the Hamiltonian
density%
\begin{equation*}
\mathbf{H}_{2}=[A^{2}+f(A^{0},A^{1})]^{1/3}
\end{equation*}%
determines an integrable hydrodynamic chain. The function $f(A^{0},A^{1})$
can be found by the method of the Hamiltonian hydrodynamic reductions.
Further details can be found in \textbf{\cite{FerMarMax1}}. Nevertheless, we
would like to emphasize that the ``asymptotic analysis'' (see the beginning
of this section) significantly simplifies computations made by the
Hamiltonian hydrodynamic reductions method.

\section{Non-Hamiltonian hydrodynamic chains}

The hydrodynamic chains associated with the Kupershmidt--Manin Poisson
brackets (see (\textbf{\ref{bhs}}) and (\textbf{\ref{aca}}))%
\begin{equation*}
A_{t^{m}}^{k}=[kA^{k+n-1}\partial _{x}+n\partial _{x}A^{k+n-1}]\frac{\delta 
\mathbf{\bar{H}}_{m+1}}{\delta A^{n}}
\end{equation*}%
possess the ``waterbag moment decomposition'' (\textbf{\ref{log}}), and the
corresponding hydrodynamic reductions are%
\begin{equation*}
a_{t}^{i}=\partial _{x}\left[ \overset{m+1}{\underset{n=0}{\sum }}(a^{i})^{n}%
\frac{\delta \mathbf{\bar{H}}_{m}}{\delta A^{n}}\right] ,
\end{equation*}%
where the Hamiltonian $\mathbf{\bar{H}}_{m+1}=\int \mathbf{H}%
_{m+1}(A^{0},A^{1},...,A^{m},A^{m+1})dx$. The hydrodynamic chains associated
with the Kupershmidt Poisson brackets (\textbf{\ref{kun}}) also possess the
waterbag moment decomposition (\textbf{\ref{kum}}), and the corresponding
hydrodynamic reductions are given by (\textbf{\ref{kuppa}}).

In the both above cases corresponding hydrodynamic chains \textit{linearly}
depend on the discrete variable $k$.

The Hamiltonian approach presented in this paper is useful for
non-Hamiltonian hydrodynamic chains too. For instance, the \textit{deformed}
Benney hydrodynamic chain (see \textbf{\cite{Fer+Dav}}, cf. (\textbf{\ref{bm}%
}))%
\begin{equation}
A_{t}^{k}=A_{x}^{k+1}+\varepsilon A^{0}A_{x}^{k}+kA^{k-1}A_{x}^{0}\text{, \
\ \ \ \ }k=0\text{, }1\text{, ...}  \label{dbm}
\end{equation}%
is not Hamiltonian, but still is integrable (at least the Hamiltonian
formalism for the above hydrodynamic chain is unknown). Nevertheless, this
hydrodynamic chain has the same moment decomposition (\textbf{\ref{log}})%
\begin{equation}
A^{n}=\frac{1}{n+1}\sum \varepsilon _{k}(u^{k})^{n+1},  \label{kof}
\end{equation}%
(where $\Sigma \varepsilon _{m}=0$, cf. (\textbf{\ref{eps}})) as the Benney
hydrodynamic chain (\textbf{\ref{bm}}), but a corresponding generating
function of conservation laws is little bit more complicated. The
corresponding hydrodynamic type system%
\begin{equation}
u_{t}^{i}=u^{i}u_{x}^{i}+\varepsilon A^{0}u_{x}^{i}+A_{x}^{0}  \label{fir}
\end{equation}%
can be written in the conservative form%
\begin{equation}
a_{t}^{i}=\partial _{x}\left[ \left( \frac{1}{\varepsilon }(\ln
a^{i}-1)+\varepsilon A^{0}\right) a^{i}\right] \text{,}  \label{secon}
\end{equation}%
where%
\begin{equation}
a^{i}=\exp (\varepsilon u^{i}).  \label{third}
\end{equation}%
Following the Hamiltonian approach, the deformed Benney hydrodynamic chain
has the generating function of conservation laws%
\begin{equation}
p_{t}=\partial _{x}\left[ \left( \frac{1}{\varepsilon }(\ln p-1)+\varepsilon
A^{0}\right) p\right] \text{,}  \label{cons}
\end{equation}%
which reduces to the first equation in (\textbf{\ref{gen}}) by $\varepsilon
\rightarrow 0$ under the substitution%
\begin{equation*}
p=\exp (\varepsilon \mu ).
\end{equation*}

\textbf{Theorem 9}: \textit{The deformed Benney hydrodynamic chain} (\textbf{%
\ref{dbm}}) \textit{satisfies the Gibbons equation (cf}. (\textbf{\ref{gib}}%
))%
\begin{equation*}
\lambda _{t}-(\frac{1}{\varepsilon }\ln p+\varepsilon A^{0})\lambda _{x}=%
\frac{\partial \lambda }{\partial p}\left( p_{t}-\partial _{x}\left[ \left( 
\frac{1}{\varepsilon }(\ln p-1)+\varepsilon A^{0}\right) p\right] \right) ,
\end{equation*}%
\textit{where the equation of the Riemann mapping is} (\textbf{\ref{ryad}}).

\textbf{Proof}: can be obtained by a straightforward substitution (\textbf{%
\ref{ryad}}) and (\textbf{\ref{dbm}}) into the above Gibbons equation.

\textbf{Corollary}: The substitution (\textbf{\ref{kof}}) in the equation of
the Riemann mapping (\textbf{\ref{ryad}}) yields again the equation of the
Riemann surface (\textbf{\ref{water}})%
\begin{equation*}
\lambda =\mu -\sum \varepsilon _{k}\ln (\mu -u^{k}).
\end{equation*}

\textbf{Remark}: Rational (Zakharov) reductions (of the Benney hydrodynamic
chain, see \textbf{\cite{Zakh}}) can be obtained by a limit from the
waterbag reductions (see, for instance, \textbf{\cite{Bogdan}}): one can
substitute the series $\tilde{u}^{(k)}=u^{k}+\upsilon ^{k}/\varepsilon
^{k}+w^{k}/(\varepsilon ^{k})^{2}+...$, $\varepsilon ^{k}\rightarrow \infty $
to the particular case of the waterbag reduction%
\begin{equation*}
\lambda =\mu -\underset{k=1}{\overset{N}{\sum }}\varepsilon _{k}\ln \frac{%
\mu -\tilde{u}^{k}}{\mu -u^{k}}.
\end{equation*}%
The same result can be obtained by a straightforward substitution the
corresponding moment decomposition (\textbf{\ref{rac}})%
\begin{equation}
A^{k}=\underset{n=1}{\overset{N}{\sum }}(u^{n})^{k}\upsilon ^{n}  \label{zak}
\end{equation}%
to the deformed Benney hydrodynamic chain. The corresponding hydrodynamic
type system (cf. the second equation below with (\textbf{\ref{fir}}))%
\begin{equation*}
\upsilon _{t}^{k}=\partial _{x}(u^{k}\upsilon ^{k})+\varepsilon
A^{0}\upsilon _{x}^{k}\text{, \ \ \ \ \ \ \ }u_{t}^{k}=\partial _{x}\left( 
\frac{(u^{k})^{2}}{2}+A^{0}\right) +\varepsilon A^{0}u_{x}^{k}
\end{equation*}%
can be written in the conservative form (can be obtained directly from (%
\textbf{\ref{cons}}) by substitution series $p^{(k)}\rightarrow
a^{k}+b^{k}\lambda +...$; cf. (\textbf{\ref{secon}}) and the first equation
from the system below)%
\begin{equation*}
a_{t}^{i}=\partial _{x}\left[ \left( \frac{1}{\varepsilon }(\ln
a^{i}-1)+\varepsilon A^{0}\right) a^{i}\right] \text{, \ \ \ \ \ \ }%
b_{t}^{i}=\partial _{x}\left[ \left( \frac{1}{\varepsilon }\ln
a^{i}+\varepsilon A^{0}\right) b^{i}\right] ,
\end{equation*}%
where $a^{i}=\exp (\varepsilon u^{i})$ (cf. (\textbf{\ref{third}})) and $%
b^{i}=\varepsilon \upsilon ^{i}\exp (\varepsilon u^{i})$.

\section{The Zakharov hydrodynamic reductions}

Let us consider the integrable hydrodynamic chain (see \textbf{\cite{Kuper}})%
\begin{equation*}
C_{t}^{k}=C_{x}^{k+1}+C^{0}C_{x}^{k}+(\alpha k+\beta )C^{k}C_{x}^{0}+\gamma
kC^{k-1}[C^{1}+\frac{\beta -\alpha -\gamma }{2}(C^{0})^{2}]_{x}\text{, \ \ \
\ \ \ }k=0,1,2,...
\end{equation*}%
This chain does not possess waterbag hydrodynamic reductions given by the
moment decomposition (\textbf{\ref{kof}}). However, the Zakharov
hydrodynamic reductions (\textbf{\ref{zak}}) exist if a hydrodynamic chain
has a \textit{linear} dependence with respect to the discrete variable $k$.
Thus, the corresponding hydrodynamic reduction is%
\begin{eqnarray*}
u_{t}^{i} &=&u^{i}u_{x}^{i}+C^{0}u_{x}^{i}+\alpha u^{i}C_{x}^{0}+\gamma
\lbrack C^{1}+\frac{\beta -\alpha -\gamma }{2}(C^{0})^{2}]_{x}\text{,} \\
&& \\
\upsilon _{t}^{i} &=&u^{i}\upsilon _{x}^{i}+\upsilon
^{i}u_{x}^{i}+C^{0}\upsilon _{x}^{i}+\beta \upsilon ^{i}C_{x}^{0},
\end{eqnarray*}%
where $C^{k}=\Sigma (u^{i})^{k}\upsilon ^{i}$. Introducing the new field
variables%
\begin{equation*}
a^{i}=(u^{i}-\gamma C^{0})^{\frac{\gamma +1}{\gamma +\alpha }},
\end{equation*}%
the \textit{first} $N$ equations can be written in the conservative form%
\begin{equation*}
a_{t}^{i}=\partial _{x}\left( \frac{\gamma +1}{2\gamma +\alpha +1}(a^{i})^{%
\frac{2\gamma +\alpha +1}{\gamma +1}}+(\gamma +1)C^{0}a^{i}\right) .
\end{equation*}%
Thus, the generating function of conservation laws%
\begin{equation}
p_{t}=\partial _{x}\left( \frac{\gamma +1}{2\gamma +\alpha +1}p^{\frac{%
2\gamma +\alpha +1}{\gamma +1}}+(\gamma +1)C^{0}p\right)   \label{a}
\end{equation}%
is obtained by the replacement $a^{i}\rightarrow p$. Substituting the Taylor
expansion $p^{(k)}\rightarrow a^{k}+\lambda b^{k}+...$, where $\lambda $ is
a local parameter, one obtains the \textit{second} $N$ conservation laws%
\begin{equation*}
b_{t}^{i}=\partial _{x}\left[ \left( (a^{i})^{\frac{\gamma +\alpha }{\gamma
+1}}+(\gamma +1)C^{0}\right) b^{i}\right] ,
\end{equation*}%
where $b^{i}=(a^{i})^{\frac{1-\beta }{\gamma +1}}\upsilon ^{i}$.

\textbf{Remark}: The Kupershmidt hydrodynamic chain (\textbf{\ref{kupu}})
(see \textbf{\cite{Kuper}}) has the generating function of conservation laws
(see (\textbf{\ref{rima}}))%
\begin{equation}
p_{t}=\partial _{x}\left( \frac{p^{\varepsilon +1}}{\varepsilon +1}+\frac{%
B^{0}}{\varepsilon }p\right) .  \label{c}
\end{equation}%
Thus, (comparing (\textbf{\ref{a}}) and (\textbf{\ref{c}})) we conclude that
both above hydrodynamic chains are equivalent under invertible
transformations $B^{k}=B^{k}(C^{0},C^{1},...,C^{k})$, $k=0,1,2,...$, where%
\begin{equation*}
\varepsilon =\frac{\gamma +\alpha }{\gamma +1}\text{, \ \ \ \ \ }\delta =%
\frac{\beta +\gamma }{\gamma +\alpha }.
\end{equation*}%
Explicit expressions can be found by the comparison two series%
\begin{equation*}
\underset{k=0}{\overset{\infty }{\sum }}\frac{B^{k}}{p^{k+1}}=(\gamma
+\alpha )\underset{k=0}{\overset{\infty }{\sum }}\frac{C^{k}}{(p+\gamma
C^{0})^{k+1}},
\end{equation*}%
where the equation of the Riemann surface is given by%
\begin{equation*}
\lambda =p^{\frac{\alpha -\beta }{\gamma +\alpha }}\left[ 1+(\alpha -\beta )%
\underset{k=0}{\overset{\infty }{\sum }}\frac{C^{k}}{(p+\gamma C^{0})^{k+1}}%
\right] .
\end{equation*}%
Then the moments $B^{k}$ can be written explicitly via the field variables $%
a^{i}$ and $b^{i}$ in the form (cf. (\textbf{\ref{rac}}))%
\begin{equation*}
B^{k}=(\gamma +\alpha )\underset{i=1}{\overset{N}{\sum }}(a^{i})^{\frac{%
(\gamma +\alpha )k+\beta -1}{\gamma +1}}b^{i}.
\end{equation*}

\section{Conclusion}

At this moment several very powerful approaches are appropriate in the
theory of classification of integrable hydrodynamic chains. These are the
symmetry approach (see \textbf{\cite{Maks+Egor}}), the method of
hydrodynamic reductions (see \textbf{\cite{Fer+Kar}}, \textbf{\cite{Gib+Tsar}%
}), the method of pseudo-potentials (see \textbf{\cite{Fer+Kar}}, \textbf{%
\cite{Zakh+multi}}), the tensor approach (\textbf{\cite{Fer+Dav}}).

However, \textit{this} Hamiltonian approach (based on the concept of the
Hamiltonian hydrodynamic reductions) in most cases allows to check
integrability and classify integrable hydrodynamic chains avoiding symbolic
software. Moreover, this method simultaneously allows to construct
infinitely many particular solutions parameterized by arbitrary constants in
the explicit form.

\section*{Acknowledgement}

I thank Maciej Blaszak, Eugeni Ferapontov, John Gibbons, Yuji Kodama, Boris
Konopelchenko, Boris Kupershmidt, Andrey Maltsev and Sergey Tsarev for their
stimulating and clarifying discussions.

I am grateful to the Institute of Mathematics in Taipei (Taiwan) where some
part of this work has been done, and especially to Jen-Hsu Chang, Jyh-Hao
Lee, Ming-Hien Tu and Derchyi Wu for fruitful discussions.

\end{document}